\documentclass[CRPHYS,Unicode,manuscript]{cedram}

\usepackage{ulem}
\usepackage{nicematrix}
\usepackage{dsfont}

\definecolor{HotCoral}{RGB}{240, 96, 92}
\definecolor{LivingCoral}{RGB}{252, 118, 106}
\definecolor{ViridianGreen}{RGB}{0, 150, 152}
\definecolor{PacificCoast}{RGB}{91, 132, 177}

\definecolor{CopenhagueRed}{RGB}{236, 125, 128}
\definecolor{CopenhagueBlue}{RGB}{168, 207, 239}
\definecolor{CopenhagueYellow}{RGB}{249, 215, 148}

\definecolor{CopenhagueGreen}{RGB}{89, 149, 152}
\definecolor{CopenhagueDarkBlue}{RGB}{43, 103, 162}

\definecolor{CopenhagueSand}{RGB}{219, 221, 208}
\definecolor{CopenhagueOrange}{RGB}{243, 170, 138}

\definecolor{CopenhagueSaturatedYellow}{RGB}{249, 199, 100}
\definecolor{CopenhagueSaturedBlue}{RGB}{118, 185, 239}

\definecolor{CopenhagueSaturatedSand}{RGB}{207, 212, 174}
\definecolor{CopenhagueSaturatedOrange}{RGB}{242, 156, 119}

\title{Dissipative parametric resonance in a modulated 1D Bose gas}

\alttitle{R\'{e}sonance param\'{e}trique dissipative dans un gaz de Bose 1D modul\'{e} }

\author{\firstname{Amaury} \lastname{Micheli}\CDRorcid{0000-0002-5240-140X}}
\address{RIKEN iTHEMS, Wako, Saitama 351-0198, Japan}
\email[A. Micheli]{amaury.micheli@riken.jp}
\author{\firstname{Scott} \lastname{Robertson}\CDRorcid{0000-0001-5919-8320}}
\address{Institut Pprime, CNRS -- Université de Poitiers -- ISAE-ENSMA, 11 Bd. Marie et Pierre Curie, 86073 Chasseneuil-du-Poitou, France}
\email[S. Robertson]{scott.james.robertson@univ-poitiers.fr}
\thanks{This work was supported by the French National Research Agency via Grant No. ANR-20-CE47-0001 associated with the project COSQUA (Cosmology and Quantum Simulation). S. R. is funded by the CNRS Chair in Physical Hydrodynamics (ref. ANR-22-CPJ2-0039-01). This work pertains to (namely, is not funded by but enters in the scientific perimeter of) the French government programs ``Investissement d'Avenir'' EUR INTREE (ref. ANR-18-EURE-0010) and LABEX INTERACTIFS (ref. ANR-11-LABX-0017-01).}


\begin{abstract}
We synthesize results of previous works to give a coherent and self-consistent account of parametric resonance in a modulated quasi-1D Bose gas in the presence of a dissipative mechanism.
The resonant behaviour is shown to be largely in line with the predictions of a phenomenological model published in 2014, while the associated dissipation rate is consistent with that derived in 2022 from three-wave mixing processes between the produced phonons and thermal fluctuations.

\end{abstract}

\begin{altabstract}
 Nous synthétisons les résultats de travaux antérieurs pour produire une présentation cohérente et auto-contenue du processus de résonance paramétrique dans un gaz de Bose quasi-1D modulé, en présence de dissipation. Nous démontrons que le déroulement du processus est en très bon accord avec les prédictions d'un modèle phénoménologique publié en 2014. De plus, nous montrons que le taux de dissipation associé est consistant avec celui dérivé en 2022 à partir de l'étude de processus de mélange à trois ondes entre les phonons produits par la modulation et les fluctuations thermiques du gaz.
\end{altabstract}

\begin{document}

\maketitle

\section{Introduction}

Quantum Field Theory suggests that the vacuum is far from empty, but rather populated by fluctuations continuously popping in and out of existence.  In particular, it makes a precise and measurable prediction that these vacuum fluctuations can be excited into real excitations by some non-trivial background. 
While Hawking radiation from a black hole is a particularly celebrated example, in which particle creation is induced by a stationary inhomogeneous background, a complementary scenario occurs when the background is homogeneous and time-dependent. In a cosmological context this leads to particle creation in the early universe~\cite{Parker,Kofman-et-al}, but it also describes laboratory-based systems such as the dynamical Casimir effect in a superconducting waveguide~\cite{Wilson,Lahteenmaki}.  It is therefore a unifying phenomenon behind many examples of Analogue Gravity that are inclined towards mimicking the early universe via some time-dependent scenario~\cite{Micheli-2023}.

One such example involves the creation of phononic quasiparticles out of vacuum in a modulated 
(quasi-)1D Bose gas~\cite{Jaskula2012}.  
In a common scenario~\cite{Jaskula2012,Butera:2022kwi}, the transverse trapping potential is varied in time, inducing a transverse motion of the Bose gas and, consequently, a time-dependence of the density and the speed of sound.   
For longitudinal excitations, this motion appears as the action of a time-varying background classical field, and quasiparticles are excited in entangled pairs of opposite momenta, just as predicted for the strong time-dependence of the early universe e.g. during inflation~\cite{Mukhanov:1982nu,Grishchuk:1990bj} and preheating~\cite{Kofman:1997yn}.

In an earlier series of papers~\cite{Busch-2014,Robertson-2017-I,Robertson-2017-II,Robertson-2018,Micheli-2022}, 
a systematic study of the expected properties of the parametric amplication induced by an oscillating transverse profile was performed.
The evolution of the number spectrum of excited quasiparticles and of the correlation amplitudes between quasiparticles of opposite momenta was described, with particular emphasis placed on the achievement of a nonseparable bipartite state for these quasiparticle pairs.  In addition, the consequences of a weak dissipation rate was studied as a possible explanation for the absence of nonseparability in the original experiment~\cite{Jaskula2012}.  In~\cite{Busch-2014}, this dissipative rate was introduced phenomenologically, while in~\cite{Robertson-2018} numerical simulations of the fully nonlinear equations showed that interactions between the quasiparticle excitations could act as an effective dissipative mechanism.  In the most recent paper~\cite{Micheli-2022}, this idea was explored in more detail in the regime of low excitation number.
It was shown that a narrow spectrum of quasiparticles would be broadened by interactions with a thermal bath of phonons, and that the initial occupation would thereby decay in time. 
We believe that~\cite{Micheli-2022} managed to pinpoint the microphysical mechanism behind the dissipation and to calculate the associated decay rate that had been introduced only phenomenologically in~\cite{Busch-2014}.

The purpose of this paper is to close the circle and show that the phenomenological study of~\cite{Busch-2014} does indeed 
describe the early-time behaviour seen in fully nonlinear numerical simulations of the 1D Bose gas, with the value of the 
dissipation rate given by the calculation recently performed in~\cite{Micheli-2022}. 
It begins in Sec.~\ref{sec:Review} with a brief review of the relevant results from earlier works.
In Sec.~\ref{sec:Dissipative_parametric_amplification} we present exact results at resonance for the dissipative analytical model introduced in~\cite{Busch-2014}.
In Sec.~\ref{sec:Numerics} we show results of fully nonlinear numerical simulations to corroborate the preceding analysis, before concluding in Sec.~\ref{sec:Conclusion}.
The appendices are devoted to a description of the error analysis applied to our numerical simulations.


\section{Review of previous results
\label{sec:Review}}

Here we review and synthesise the relevant findings of the theoretical works~\cite{Busch-2014,Robertson-2017-I,Robertson-2017-II,Robertson-2018,Micheli-2022}.
We start by describing the general class of bipartite Gaussian homogeneous states~\cite{Campo-Parentani-2005} and the characterisation of their entanglement properties.  
We then describe how the modulation of an elongated Bose gas predicts parametric amplification of pairs of entangled quasiparticles in just such a state~\cite{Busch-2014}.
Finally, the inclusion of weak dissipation in our theoretical model (as developed in 
~\cite{Busch-2014}) is reviewed, as is the microphysical origin of this dissipation described in~\cite{Micheli-2022}.

\subsection{Characterisation of bipartite quasiparticle states}

Consider a linear evolution of a quantum field in a time-varying background that remains always homogeneous.
Certain symmetries of the evolution ensure that certain properties of the state are preserved.
Invariance under spatial translations and parity inversions ensures that the state remains statistically homogeneous and isotropic. 
Linearity of the evolution ensures that an initially Gaussian state remains Gaussian, and in conjunction with homogeneity, ensures that each bipartite sector $(k,-k)$ evolves independently. 
In the absence of any initial correlations, the full state of the field can at any time be written as a product over all bipartite states
\begin{equation}
    \hat{\rho} = \bigotimes_{k} \hat{\rho}_{k,-k} \,.
\end{equation}
We thus restrict our attention to a single bipartite state $\hat{\rho}_{k,-k}$, with the individual modes having quantum amplitudes $\hat{b}_{k}$ and $\hat{b}_{-k}$ obeying the bosonic commutation relations $\left[ \hat{b}_{\pm k} \,,\, \hat{b}_{\pm k}^{\dagger} \right] = 1$.

Since the preserved properties of Gaussianity, isotropy, and homogeneity are all properties of a thermal state, an initial thermal state will evolve into a state which, though no longer thermal, will still be Gaussian, isotropic, and statistically homogeneous~\cite{Campo-Parentani-2005}.  
A bipartite state with such properties 
is completely characterised by only two expectation values (corresponding to three real parameters)
\begin{equation}
n_{k} = \Big\langle \hat{b}_{k}^{\dagger} \hat{b}_{k} \Big\rangle \,, \qquad c_{k} = \Big\langle \hat{b}_{k} \hat{b}_{-k} \Big\rangle \,.
\label{eq:n_and_c}
\end{equation}
The first, $n_{k}$, is just the mode occupation number (with $n_k = n_{-k}$ by isotropy).  The second, $c_{k}$, is the pair-correlation amplitude between modes $\pm k$ and is in general complex. 

The question of entanglement between modes $k$ and $-k$ naturally arises. 
There exist several inequivalent criteria of entanglement. 
In this paper (as in~\cite{Busch-2014,Robertson-2017-I,Robertson-2018}) we shall principally adopt the notion of {\it nonseparability} as our criterion of choice.
A bipartite state is said to be {\it separable}~\cite{Werner-1989} when it can be written in the form
\begin{equation}
    \hat{\rho}_{k,-k} = \sum_{j} P_{j} \, \hat{\rho}_{k}^{(j)} \otimes \hat{\rho}_{-k}^{(j)} \,,
\end{equation}
where $P_{j} \geq 0$ for all $j$.  The $P_{j}$ can then be identified with a classical probability distribution, and the correlations interpreted classically.  Whenever the bipartite state cannot be written in this form, we refer to it as 
{\it nonseparable}.
Remarkably, the expectation values $n_{k}$ and $c_{k}$ yield a simple sufficient condition for nonseparability (which becomes 
an equivalence under the assumptions of Gaussianity, isotropy and homogeneity)~\cite{Campo-Parentani-2005,Busch-Parentani-2013}: the state must be nonseparable whenever
\begin{equation}
    \Delta_{k} = n_{k} - \left|c_{k}\right| < 0 \,.
    \label{eq:nonsep}
\end{equation}
Moreover, since quantum mechanics provides an upper bound on $\lvert c_{k} \rvert$ of the form
\begin{equation}
\label{eq:max_c}
    \lvert c_{k} \rvert^{2} \leq n_{k} \left(n_{k} + 1\right) \,,
\end{equation}
we find that $\Delta_{k} > -1/2$.  There is thus a rather small window of $\Delta_{k}$ where we have a clear indication of 
the nonseparability of the state.

\subsection{Evolution of phononic quasiparticles in a modulated Bose gas}
\label{subsec:modulation_no_dissipation}

In~\cite{Busch-2014}, the effect of a time-dependent background on the dynamics of quasiparticles in a homogeneous 1D Bose gas is considered.
Although the analysis is fairly generic, to fix ideas it will be useful to recall the main steps here, rewritten in a form more amenable to the analysis of later papers.
We recall the steps quickly, and we refer to~\cite{Micheli-2023} for a detailed description.

The 1D atomic Bose gas, of fixed atom number, can for our purposes be sufficiently described by the following Hamiltonian
\begin{equation}
    \hat{H} = \int_{0}^{L} dx \, \left\{ \frac{\hbar^{2}}{2m} \partial_{x}\hat{\Psi}^{\dagger} \partial_{x}\hat{\Psi} + \frac{g}{2} \hat{\Psi}^{\dagger 2} \hat{\Psi}^{2} \right\} \,,
    \label{eq:full_Hamiltonian}
\end{equation}
where $\hat{\Psi}$ is the atomic field operator, $m$ is the atomic mass, $L$ is the length of the gas (considered as periodic), and $g$ is the effective 1D interaction constant that governs the strength of two-body contact interactions.  This interaction constant depends on the degree of confinement in the transverse direction, with greater confinement yielding stronger interactions in the dimensionally reduced system~\footnote{Explicitly, we have $g = g_{3D} G (\rho_0 a_s)/ 2 \pi \sigma^{2}$~\cite{Robertson-2017-I,Robertson-2018}, where $g_{3D}$ is the 3D interaction strength characterising two-body contact interactions, $a_s = m g_{3D} / 4 \pi \hbar^2$ is the $s$-wave scattering length, $\rho_0$ is the linear density of the gas, $G (\rho_0 a_s)$ is an adimensional function that is typically of order $O(1)$, and $\sigma$ is the transverse width of the gas. \label{fn:3d_to_1d}}. Inducing a transverse modulation of the gas is thus equivalent, in the 1D description, to a modulation of the 1D interaction strength $g$.

Given the quasicondensate nature of the 1D Bose gas~\cite{Petrov-et-al-2004}, it is appropriate to perform a Madelung transformation $\hat{\Psi} = e^{i\hat{\theta}} \sqrt{\hat{\rho}}$, where $\hat{\rho}$ and $\hat{\theta}$ are the density and phase of the gas~\cite{Mora-Castin}.
They are assumed to obey the commutation relation
\begin{equation}
\label{eq:commutator_rho_theta}
    \left[ \hat{\rho} \left( x \right) , i \hat{\theta} \left( x^{\prime} \right) \right] = \delta \left( x - x^{\prime} \right) \, ,
\end{equation}
and the Hamiltonian is taken to be~\footnote{Note that although the Hermitian operator $\hat{\theta}$ is actually ill-defined, meaningful  predictions can be derived in the quasi-condensed regime considered here~\cite{Mora-Castin}. The quantum theory for the density and phase operators is obtained by first performing the transformation $(\Psi,\Psi^{\star}) \to (\rho , i \theta)$ at the classical level, where it can be shown to be canonical, and then promoting these variables to operators obeying the canonical commutation relations Eq.~\eqref{eq:commutator_rho_theta} and picking the ordering~\eqref{eq:full_Hamiltonian_madelung} for the Hamiltonian. }
\begin{equation}
\label{eq:full_Hamiltonian_madelung}
\hat{H} =  \int_{0}^{L} \left[ \frac{\hbar^2}{8m \hat{\rho}} \left( \frac{\partial \hat{\rho} }{\partial x} \right)^2 + \frac{\hbar^2}{2m}  \frac{\partial \hat{\theta} }{\partial x} \hat{\rho}  \frac{\partial \hat{\theta} }{\partial x} + \frac{g}{2} \hat{\rho}^2 \right] \mathrm{d}x \, .
\end{equation}

A quasi-condensed state is characterised by small phase gradients $\partial_x \hat{\theta}$ and small density fluctuations $\delta \hat{\rho}$ about $\hat{\rho}_0$ (the $k=0$ component of the density). Using the Heisenberg equations one can show that $\hat{\rho}_0$ is exactly conserved, a consequence of the conservation of atom number. As in the standard Bogoliubov procedure for a true condensate, we then treat the large background density as a c-number $\hat{\rho}_0 \approx \rho_0 \hat{\mathds{1}}$ and define the $k \neq 0$ components as a perturbation
\begin{equation}
    \delta\hat{\rho}(x) = \sqrt{\frac{\rho_{0}}{L}} \sum_{k \neq 0} \delta\hat{\rho}_{k} e^{i k x} \, .
\end{equation}
Only gradients of the phase appear in the Hamiltonian~\eqref{eq:full_Hamiltonian_madelung} so that its $k=0$ average value $\hat{\theta}_0$ is irrelevant to the dynamics~\footnote{Solving the equations of motion at first order in $\rho_0$ gives the running phase $\hat{\theta}_0 \approx - g \rho_0 / \hbar \hat{\mathds{1}}$. }. We define
\begin{equation}
    \delta\hat{\theta}(x) = \frac{1}{\sqrt{\rho_{0}L}} \sum_{k \neq 0} \delta\hat{\theta}_{k} e^{i k x} \, ,
\end{equation}
whose gradients are small.
The Hamiltonian can be expanded to arbitrarily high orders in the perturbation fields
$\hat{H} = E^{(0)} + \hat{H}^{(2)} + \hat{H}^{(3)} + \hat{H}^{(4)} + \ldots$ (where $E^{(0)}$ is the energy associated to the $k=0$ components).  Here, it suffices to consider only the quadratic part of the Hamiltonian
\begin{equation}
    \hat{H}^{(2)} = \sum_{k \neq 0} \left[ \frac{\hbar^{2} k^{2}}{2m} \delta\hat{\theta}_{k} \delta\hat{\theta}_{-k} + \left( \frac{\hbar^{2}k^{2}}{8m} + \frac{g \rho_{0}}{2} \right) \delta\hat{\rho}_{k} \delta\hat{\rho}_{-k} \right] \,.
\end{equation}
The quasiparticles of the system are those excitations -- linear combinations of the $\delta\hat{\rho}_{k}$ and $\delta\hat{\theta}_{k}$ -- that diagonalise $\hat{H}^{(2)}$, which then describes a collection of non-interacting quasiparticles
\begin{equation}
    \hat{H}^{(2)} = \sum_{k} \hbar \omega_{k} \left( \hat{b}_{k}^{\dagger} \hat{b}_{k} + \frac{1}{2} \right) \,.
    \label{eq:hamiltonian_quadratic}
\end{equation}
The $\omega_{k}$ are the associated quasiparticle frequencies
\begin{eqnarray}
    \omega_{k}^{2} &=& \frac{g \rho_{0}}{m} k^{2} + \left(\frac{\hbar k^{2}}{2m}\right)^{2} \nonumber \\
    &=& c^{2} k^{2} \left( 1 + \frac{1}{4} k^{2}\xi^{2} \right) \,,
    \label{eq:frequency}
\end{eqnarray}
where $c = \sqrt{g\rho_{0}/m}$ is the low-$k$ wave speed and $\xi = \hbar/mc$ is the healing length.  We also define the ``healing time'' $t_{\xi} = \xi/c = \hbar/mc^{2}$, which allows us to write the frequency in fully adimensional form
\begin{equation}
    \left(\omega_{k} t_{\xi} \right)^{2} = \left(k\xi\right)^{2} + \frac{1}{4} \left(k\xi\right)^{4} \,.
\end{equation}
Stopping at quadratic order $\hat{H}^{(2)}$ in the expansion of the Hamiltonian amounts to considering free ({\it i.e.}, non-interacting) quasiparticles, and corresponds to the Bogoliubov-de Gennes (BdG) approximation.

Let us now suppose that the condensate is modulated in such a way that the 1D interaction strength $g$ varies in time according to
\begin{equation}
\label{eq:g_modulation}
    g(t) = g \left[ 1 + a \, {\rm sin}\left( \omega_{p} t \right) \right] \,.
\end{equation}
As mentioned above, this can be achieved by varying the transverse trapping potential so as to induce a transverse oscillation of the gas.
From Eq.~(\ref{eq:frequency}), the quasiparticle frequencies are modulated according to 
\begin{equation}
    \omega_{k}^{2}(t) = \omega_{0}^{2} \left[ 1 + A_{k} \, {\rm sin}\left( \omega_{p} t \right) \right] \,.
\label{eq:frequency_modulation}
\end{equation}
Here, $\omega_{0}^{2}$ is the time-averaged value of $\omega_{k}^{2}$, $\omega_{p}$ is the modulation frequency, and $A_{k}$ is the relative amplitude of the modulation which has the following $k$-dependence
\begin{equation}
\label{eq:effective_amplitude_mod}
    A_{k} = \frac{a}{1+k^{2}\xi^{2}/4} \,.
\end{equation}
The variation of $g$ induces a variation in the definition of the quasiparticle amplitudes $\hat{b}_{k}$ (through the coefficients relating them to the density and phase perturbations), which standardly leads to quasiparticles production~\cite{Parker}.

It was shown in~\cite{Busch-2014} that if $\omega_{0}$ lies in a window of width $A_k \, \omega_{p}/4$ centered around $\omega_{p}/2$ ({\it i.e.}, if $\omega_{0} \in \left[ \omega_{p}/2 \times \left(1-A_k/4\right) \,, \omega_{p}/2 \times \left(1+A_k/4\right) \right]$), then $n_{k}$ and $| c_{k} |$ grow exponentially in time.  Note that since $A_{k}$ vanishes in the limit of large $k$, high-frequency modes are essentially unaffected by the modulation. 
This exponential growth within a window centred at half the driving frequency is standard in parametric resonance, and the behaviour can be understood in terms of a Mathieu equation~\cite{mathieuMemoireMouvementVibratoire1868,kovacicMathieuEquationIts2018} on the density perturbations~\cite{Micheli-2023}.
Since isotropy implies $\omega_{k}=\omega_{-k}$, we get pairs of resonant modes $\pm k$ rather than single ones, and $|c_k|$ encodes the correlation of these co-growing pairs.
In particular, for the exactly resonant mode ($\omega_{0} = \omega_{p}/2$), we have~\cite{Busch-2014,Micheli-2022}
\begin{eqnarray}
n_{k}(t) + \frac{1}{2} &=& \left(n_{k}^{\rm in} + \frac{1}{2}\right)\left[1 + 2 \, {\rm sinh}^{2}\left( \frac{1}{2} G_k t \right) \right] \, 
\label{eq:sol_nk_no_dissipation} \\
\lvert c_{k}(t) \rvert &=&  \left( n_{k}^{\rm in} + \frac{1}{2} \right) {\rm sinh}\left(  G_k t \right) \, , 
\label{eq:sol_ck_no_dissipation}
\end{eqnarray}
where $G_k = A_k \omega_{p}/4$.  
After an initial period where $n_{k}$ and $|c_k |$ grow as $t^{2}$ and $t$, respectively, the growth becomes exponential 
with $n_{k} \sim \left|c_{k}\right| \sim e^{G_{k}t}$, so that $G_{k}$ corresponds to the growth rate~\footnote{\label{fn:R} For frequencies inside the resonant window but not exactly at resonance, the growth rate falls off as $\sqrt{1-R_k^{2}}$ where $R_k = 4 ( 2\omega_0 - \omega_p )/ A_k \omega_p$ is a dimensionless detuning parameter that goes to $\pm 1$ at the edges of the resonant window~\cite{Busch-2014}.}.  Notice that $\Delta_k = n_k - \lvert c_{k} \rvert$ decays exponentially fast to $-1/2$, {\it i.e.}, the state {\it always} becomes nonseparable, no matter the initial temperature or the growth rate, as long as we let the modulation run for a sufficiently long time.

\subsection{Effective dissipation of quasiparticles
\label{subsec:quasiparticle_decay}}

A brief section of~\cite{Busch-2014} was dedicated to weak dissipative effects.  The dissipation was introduced phenomenologically by inventing a rate $\Gamma_{k}$ and assuming that this acts continuously on both $n_k$ and $c_k$ throughout their evolution~\footnote{To match the conventions of~\cite{Micheli-2022}, we define $\Gamma_k$ as the decay rate of $n_k$ and $c_k$.  Note the difference with respect to~\cite{Busch-2014} where $\Gamma_{k}$ referred to the decay rate of the Bogoliubov coefficients $\alpha_{k}$ and $\beta_{k}$, so that $n_{k}$ and $c_{k}$, being related to the square of the Bogoliubov coefficients, decayed at rate $2 \Gamma_k$.}.
Thinking of the modes $\pm k$ as a collection of correlated particles, it is intuitive that when a particle in either of the modes is lost, and $n_k$ decreases, then so does the correlation and $c_k$ decreases. However, $n_k$ was assumed to decay not to zero but to an equilibrium value $n_{\rm eq}$, which was identified with the thermal population. The correlation amplitude $c_k$ was assumed to have no equilibrium value and to decay to zero. 
Under these assumptions, effective equations of motion were derived (see Appendix~B of~\cite{Busch-2014}).
We will return to these equations and their solutions in more detail in Sec.~\ref{sec:Dissipative_parametric_amplification}, where we compare their predictions with the results of fully nonlinear numerical simulations.

In~\cite{Robertson-2018}, it was numerically observed in simulations of the nonlinear equations of motion associated with the full Hamiltonian~(\ref{eq:full_Hamiltonian}) that phonon-phonon interactions can induce an effective dissipation that would act against the parametric resonance.  In this case, the effective dissipation was nonlinear in the occupation number, coming into play at relatively late times when the occupation number of the resonant modes is a significant fraction of the total atom number (and leading, among other things, to a significant degree of second harmonic generation: $k + k \to 2k$).  This is {\it not} the regime described by the dissipative model of~\cite{Busch-2014}, where the decay rate is independent of $n_k$ and $c_k$.  However, it does provide a proof of concept concerning the origins of the dissipative mechanism in the interactions between quasiparticles.

\begin{figure}
    \centering
\centering
\includegraphics[height=0.2\textheight]{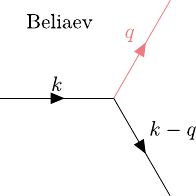}
\qquad
\qquad
\includegraphics[height=0.2\textheight]{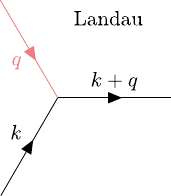}
\caption{Diagrammatic representation of the Beliaev (left) and Landau (right) three-body phonon scattering processes. Note that the arrows do not represent charge but momentum direction, whose magnitude is given above the edges. The arrows corresponding to phonons in the thermal part of the spectrum ($q \approx 0$) are shown in red.
\label{fig:landau_Belaiev}}
\end{figure}

For quasiparticle interactions to induce an effective {\it linear} decay rate ({\it i.e.}, independent of $n$ and $c$, like that considered in~\cite{Busch-2014}), we require a large ``background'' of quasiparticles that is separate from those involved in the parametric resonance.  This is provided by a thermal bath, and it is precisely the effective dissipation induced by this thermal bath in a 1D quasicondensate that is the study of~\cite{Micheli-2022}.   Since $n_{q}^{\rm th} \sim 1/\left|q\right|$ for momentum $q$ very small, the occupation numbers of the IR modes are sufficiently large that the fluctuation at $k$ is sensitive to their presence, and the effect of the corresponding phonon-phonon interactions on the mode $k$ cannot be neglected.  At lowest order, these interactions correspond to three-wave mixing processes (encoded in the third-order Hamiltonian $\hat{H}^{(3)}$): either a fluctuation at $k$ combines with one at $q$ (a small thermal wave number) to produce a fluctuation at $k+q$, a so-called Landau process; or a fluctuation at $k$ is (through a process akin to stimulated emission) encouraged to split into fluctuations at $k-q$ and $q$ by the already large occupation of the mode $q$, a so-called Beliaev process.
There is thus a tendency for the excitations in mode $k$ to be kicked into neighbouring modes, and focusing on mode $k$ itself, we see that this action of the thermal bath behaves like a dissipative mechanism.
The associated decay rate was extracted using a version of the Fermi Golden Rule applicable to this situation of broadening.  The result is
\begin{equation}
    \Gamma_k = \frac{k_{B} T}{\hbar} \, \frac{1}{\rho_{0} \xi} \, f\left(k \xi\right) \,.
    \label{eq:Gamma_formula}
\end{equation}
Here, $\rho_{0}$ is the one-dimensional atom number density of the gas, while $\xi$ is the healing length.  $f(z)$ is a smooth function that is around $3$ for $z \leq 1$ and is approximately equal to $z$ for $z \geq 1$.
Prediction~(\ref{eq:Gamma_formula}) for the decay rate was checked against fully nonlinear numerical simulations in~\cite{Micheli-2022}, though only its effect on $n_k$ was considered.

Deviations from prediction~(\ref{eq:Gamma_formula}) were identified and studied in Appendix~B of~\cite{Micheli-2022}, in the context of the relaxation of an initially excited mode in an otherwise stationary quasicondensate. First, the decay only becomes exponential for times longer than a $k$-dependent critical time $t_{\mathrm{crit}}$. Second, at late times, finite-size effects tend to slow down the decay, and can suppress it altogether for a gas whose length $L$ is smaller than the quasicondensate coherence length $r_0$~\cite{Mora-Castin}. Finally, the prediction of exponential decay is obtained neglecting the contribution of inverse processes from nearby modes, which also slow the decay. These may become important when a large number of excitations is produced, {\it e.g.}, after an extended period of parametric resonance.


\section{Parametric amplification in presence of quasiparticle decay
\label{sec:Dissipative_parametric_amplification}}

In this section, we dig into the details of the phenomenological dissipative model developed in~\cite{Busch-2014}.  
We consider analytical solutions (which were absent in~\cite{Busch-2014}) for the evolution at resonance,
using them to give an in-depth description of the interplay between growth and dissipation in the regimes of exponential and saturated growth.
For now the dissipative rate $\Gamma$ is left arbitrary, and we do not refer to or rely on the microphysical origins of the decay.
In Sec.~\ref{sec:Numerics}, dedicated to numerical simulations, we will be able to assess the validity of the general behaviour described here when the decay rate  is fixed to that of prediction~(\ref{eq:Gamma_formula}).

\subsection{An analytic solution at resonance}

In~\cite{Busch-2014}, assuming a constant decay rate $\Gamma$ that acts continuously on both $n$ and $c$ (we now drop the explicit dependence on $k$ for ease of notation), the following equations of motion were derived\footnote{Note that, 
we have implictly factored out the free evolution of $c$ corresponding to a running phase with angular velocity $\sim 2 \omega$, so that at exact resonance $c$ is purely real.  See~\cite{Busch-2014} for more details.}
\begin{equation}
\label{eq:eomncdissipapprox}
\begin{split}
(\partial_t +  \Gamma)\left(n-n_{\rm eq}\right) &=  G \, \Re \left[ e^{-i R G t} c \right] , \\
(\partial_t +  \Gamma)c &= G \, e^{i R G t } \left(n + \frac{1}{2} \right) ,
\end{split}
\end{equation}
where $G = A \omega_{p}/4$ is the would-be growth rate (in the absence of dissipation), $R$ is the dimensionless detuning from resonance (see footnote~\ref{fn:R}), and $n_{\rm eq}$ is an equilibrium occupation number that should be related to the same environment coupling responsible for the dissipation.  As in~\cite{Busch-2014}, we will typically identify $n_{\rm eq}$ with $n_{\rm th}$, the thermal bath of phonons at a given temperature $T$. 

We first notice that, in the absence of driving ({\it i.e.}, $G=0$), Eqs.~\eqref{eq:eomncdissipapprox} yield a linear exponential decay of $n-n_{\rm eq}$ and $c$, both at the rate $\Gamma$.  Though not studied systematically, a comparison of the decay for different initial values of $n$ (differing by an order of magnitude) was done in~\cite{Micheli-2022} (see Fig.~1 of~\cite{Micheli-2022}) using fully nonlinear TWA simulations.  The comparison strongly suggests that the decay is indeed linear, in accordance with Eqs.~\eqref{eq:eomncdissipapprox}~\footnote{The observant reader will note that, in Fig.~1 of~\cite{Micheli-2022}, the asymptotic value $n_{\rm eq}$ seems to vary more or less linearly with the initial occupation number, in contradiction with the assumption made in~\cite{Busch-2014} that $n_{\rm eq}$ be identified with the thermal population $n_{\rm th}$ and therefore independent of the number of phonons initially injected.   
However, we believe this scaling of the effective $n_{\rm eq}$ to be related to the fact that $n$ decays by broadening, the total number of phonons within a narrow spectral window being conserved (as was also checked numerically in~\cite{Micheli-2022}).  We expect that, in the continuum limit where there are an infinity of nearby modes to absorb the decaying phonons, $n_{\rm eq}$ should indeed become $n_{\rm th}$.}.

Let us now allow $G \neq 0$ but restrict our attention to the case of exact resonance, $R=0$.
In that case Eqs.~\eqref{eq:eomncdissipapprox} can be solved explicitly by taking their sum and difference to arrive at the following decoupled equations
\begin{eqnarray}
\label{eq:eom_n_c_decoupled}
    \left(\partial_{\tau} + \alpha - 1 \right) \left(n+\frac{1}{2}+c\right) &=& \alpha \left(n_{\rm eq} + \frac{1}{2}\right) \,, \nonumber \\
    \left(\partial_{\tau} + \alpha + 1 \right) \left(n+\frac{1}{2}-c\right) &=& \alpha \left(n_{\rm eq}+\frac{1}{2}\right) \,.
\end{eqnarray}
where we have defined the dimensionless quantities $\alpha = \Gamma/G$ and $\tau = G t$.
Imposing that the initial state is in thermal equilibrium with $n(0)=n_{\rm eq}=n_{\rm th}$ and $c(0)=0$, the solution is
\begin{eqnarray}
\label{eq:sol_nk_ck_exact_resonance}
n+\frac{1}{2}+c &=& \left(n_{\rm th}+\frac{1}{2}\right) \frac{e^{(1-\alpha)\tau} -\alpha}{1-\alpha} \, , \nonumber \\
    n+\frac{1}{2}-c &=& \left(n_{\rm th}+\frac{1}{2}\right) \frac{e^{-(1+\alpha)\tau} + \alpha}{1+\alpha} \, ,
\end{eqnarray}
whereupon we can again take the half-sum and half-difference to arrive at explicit expressions for $n$ and $c$
\begin{eqnarray}
\label{eq:n_c_initial_thermal}
n+\frac{1}{2} &=& \frac{n_{\mathrm{th}}+\frac{1}{2}}{1-\alpha^2} \left\{ -\alpha^{2} + e^{- \alpha \tau} \left[ \mathrm{cosh} (\tau) + \alpha \, \mathrm{sinh} (\tau)  \right] \right\} \,, \label{eq:n_res} \\
c &=& \frac{n_{\mathrm{th}}+\frac{1}{2}}{1-\alpha^2} \left\{ - \alpha + e^{- \alpha \tau} \left[ \mathrm{sinh} (\tau) + \alpha \, \mathrm{cosh} (\tau)  \right] \right\} \, . \label{eq:c_res}
\end{eqnarray}
The behaviour at resonance is thus fully characterised by the two dimensionless numbers $n_{\rm th}$ and $\alpha$.  The only missing ingredient is the time scale, determined by $G$.

The analytic predictions of Eqs.~\eqref{eq:n_res} and~(\ref{eq:c_res}) represents an improvement over our previous studies.  In particular, in~\cite{Micheli-2022} the behaviour of $n(t)$ during parametric amplification was fitted to the following template
\begin{equation}
    n_{\rm fit}(t) + \frac{1}{2} = \left(n_{\rm th}+\frac{1}{2}\right) \left\{ 1 + 2 \, {\rm sinh}^{2}\left[\frac{1}{2}\left(G-\Gamma\right)t\right] \right\} \,.
    \label{eq:old_n_template}
\end{equation}
This is a generalization of the non-dissipative result~(\ref{eq:sol_nk_no_dissipation}) with $n_{k}^{\rm in} \to n_{\rm th}$ and the reduction of the growth rate $G \to G-\Gamma$.  This template was used in~\cite{Micheli-2022} to extract best-fit values for the dissipative rate $\Gamma$.  However, Eq.~\eqref{eq:n_res} is {\it not} equivalent to this.  Discrepancies show up quickly, in the early-time behaviour: whereas Eq.~\eqref{eq:old_n_template} clearly predicts no evolution of $n$ at all when the reduced growth rate $G-\Gamma$ vanishes, Eqs.~\eqref{eq:n_res} and~(\ref{eq:c_res}) give, at early time,
\begin{equation}
    n + \frac{1}{2} = \left(n_{\rm th}+\frac{1}{2}\right) \left( 1 + \frac{1}{2} \tau^{2} + \ldots \right) \,, \qquad c = \left(n_{\rm th}+\frac{1}{2}\right) \left( \tau - \frac{1}{2} \alpha \tau^{2} + \ldots \right) \,.
\label{eq:n_c_early-time}
\end{equation} 
These show, in particular, that $n$ experiences an initial stage of quadratic growth {\it irrespective} of the value of $\alpha$.
Even at late times we find discrepancies between predictions~\eqref{eq:n_res} and~\eqref{eq:old_n_template}, for Eqs.~\eqref{eq:n_res} and~(\ref{eq:c_res}) yield the late-time behaviour
\begin{equation}
    n+\frac{1}{2} \sim c \sim \left(n_{\rm th}+\frac{1}{2}\right) \frac{e^{\left(1-\alpha\right)\tau}}{2\left(1-\alpha\right)} \,.
\end{equation}
While this does indeed correspond to exponential growth of $n$ at the reduced rate $G-\Gamma$ (as long as $\alpha = \Gamma/G < 1$), it comes with a coefficient of $1/(1-\alpha)$ that is not reproduced by Eq.~\eqref{eq:old_n_template}. 
Given these discrepancies, we might expect the results of~\cite{Micheli-2022} to have been skewed a little by the use of a different analytical model from the one predicted here.  A key result of the current paper is that Eq.~\eqref{eq:n_res} indeed provides a much better description of the observed behaviour of $n(t)$ than the naive model~\eqref{eq:old_n_template} adopted in~\cite{Micheli-2022}.

\subsection{Asymptotic behaviours}
\label{subsec:late_time}

Focusing on the late-time behaviour, we may immediately make two important observations that were already highlighted in~\cite{Busch-2014}.

\subsubsection{Growth and saturation}
\label{subsubsec:growth_and_saturation}

The first observation concerns the competition between growth and decay that leads either to continued growth or to saturation.
The late-time evolution of $n$ and $c$ is dominated by the first of Eqs.~\eqref{eq:sol_nk_ck_exact_resonance}, which shows that there are two distinct regimes.
On the one hand, when the growth rate is larger than the decay rate ($\alpha < 1$), $n$ and $c$ grow exponentially at the reduced growth rate $G - \Gamma = G\left(1-\alpha\right)$. 
On the other hand, when the decay rate is larger than the growth rate ($\alpha > 1$), we still have a production of correlated quasiparticles since $n$ and $c$ both grow in time, but now they saturate at finite asymptotic values. We refer to this regime as saturated growth. The critical case $\alpha =1$ gives an asympotically linear growth of $n$ and $c$.

It is natural to wonder why $n$ and $c$ grow at all when the dissipation rate exceeds the growth rate. 
In answer to this, we recall Eqs.~\eqref{eq:eomncdissipapprox} with $R=0$, paying particular attention to the source terms on the right-hand side that are ``switched on'' by the growth rate $G$.  Even when $n-n_{\rm eq}$ and $c$ are initially zero, the growth of $c$ is sourced by $G$ times $n+1/2$, which thanks to quantum vacuum fluctuations is always non-zero\footnote{Classically it would only be $n$ that appears here, and an initial occupation would be required to seed the growth of $c$.  However, quantum mechanics adds a contribution from vacuum fluctuations (encoded in the extra $1/2$), which leads to growth of $c$ even if the initial state is vacuum.}.

In turn, the non-zero $c$ appears as a source term on the first of Eqs.~\eqref{eq:eomncdissipapprox}, engendering a growth of $n-n_{\rm eq}$. 
Hence there is always some growth of $n$ and $c$, no matter the value of $\Gamma$; and moreover, the initial growth rates are in fact $\Gamma$-independent, in accordance with the early-time expansions~(\ref{eq:n_c_early-time}). On the other hand, note that the amplitude of a parametric oscillator with viscous damping at rate $2 \Gamma$ would only grow when $\Gamma > G$~\cite{kovacicMathieuEquationIts2018}; the growth is then exponential at rate $G - \Gamma$ in line with our old template~\eqref{eq:old_n_template}. One can check that, neglecting the source terms $n_{\mathrm{eq}}$ and $1/2$ (which corresponds to setting the right-hand sides of Eqs.~(\ref{eq:eom_n_c_decoupled}) to zero), we recover such damped exponential growth for $n$ and $c$. The presence of the source terms is physically necessary as they account for the noise that the environment necessarily generates in the system along with dissipation.

\subsubsection{Nonseparability}

The second point to note concerns the behaviour of $n+1/2-c = \Delta + 1/2$, which describes the evolution of the nonseparability of the state. At resonance, $\Delta$ decreases monotonically 
in time, converging to the asymptotic value 
\begin{equation}
\label{eq:asymptotic_Delta}
\Delta_{\infty} = - \frac{1}{2} \times \frac{1 - 2 \alpha n_{\mathrm{eq}}}{\alpha + 1} \, ,
\end{equation}
which is \textit{positive} when $2 \alpha n_{\mathrm{eq}} > 1$.  That is, if the product of the dissipation rate and the equilibrium population is sufficiently large, then we never reach a nonseparable state.
This threshold can be understood by referring to the second of Eqs.~\eqref{eq:eom_n_c_decoupled}, which after some rearranging and replacing of dimensionful quantities, can be written in the form
\begin{equation}
    \left( \partial_{t} + \Gamma + G \right) \left( n - c \right) = \Gamma n_{\rm eq} - \frac{G}{2} \,.
    \label{eq:Delta_evolution}
\end{equation}
Note that the source term here is constant, independent of $n$ and $c$.
The $\Gamma n_{\rm eq}$ term encodes the effect of the dissipative mechanism. 
The $G/2$ term can be traced back to the commutator $[\hat{b}_{k},\hat{b}_{k}^{\dagger}] =1$ and therefore encodes the contribution from vacuum fluctuations to the dynamics; neglecting this term would correspond to assuming purely classical dynamics.
We thus see immediately that, in the classical non-dissipative case, $\Delta = n-c$ is exactly conserved.
The positivity of the dissipative source term means that it tends to degrade the degree of correlation, while the negativity of the quantum contribution shows that it tends to increase the degree of correlation.  This negative source term is strictly non-classical, and necessary for an initially positive $\Delta$ to become negative; that is, the production of an entangled state implies at least partial seeding by vacuum fluctuations, justifying the interpretation of 
entanglement as a signature of a genuine quantum effect.
The simultaneous presence of the two source terms on the right-hand side of Eq.~(\ref{eq:Delta_evolution}) sets up a competition between dissipative effects and sourcing by vacuum fluctuations.  Which of these is larger dictates whether the final state is separable or entangled.

 \subsubsection{Independence of thresholds}

Interestingly, the threshold delimiting growth from saturation ($G=\Gamma$) is distinct from the threshold delimiting separability and nonseparability of the final state ($G=2n_{\rm eq} \Gamma$).  Any combination of asymptotic behaviours is therefore possible, depending on the regime.  In particular, if $n_{\rm eq} > 1/2$, then we may have continued exponential growth of $n$ and $c$ while the state remains always separable.  By contrast, when $n_{\rm eq} < 1/2$, there exists a regime where $n$ and $c$ saturate, yet the final state is nonseparable.


\section{Numerical observations
\label{sec:Numerics}}

We now turn to numerical simulations of the fully nonlinear dynamics of a modulated 1D quasicondensate.
The goal of these simulations is twofold: to test whether Eqs.~\eqref{eq:n_res} and~\eqref{eq:c_res} provide a good description of the dynamics at resonance and correctly predict the existence of different growth regimes as described above; and to validate the prediction~(\ref{eq:Gamma_formula}) for the associated dissipative rate, particularly as it applies to the early stages of parametric amplification. 
In this way, we correct for a bias in the results of~\cite{Micheli-2022} (due to the use of too naive a template for $n(t)$, see the discussion after Eq.~(\ref{eq:old_n_template})); we synthesise the results of~\cite{Busch-2014} and~\cite{Micheli-2022} through the determination of the appropriate dissipative rate, which had been left arbitrary in~\cite{Busch-2014}; and we extend the numerical observations of~\cite{Robertson-2018} to the early stages of the parametric resonance, where the number of produced phonons is still relatively small.

We simulate\footnote{For more detailed information on our simulations, we refer the reader to~\cite{Micheli-2022} and Sec.~3.5 of~\cite{Micheli-2023}.} the parametric growth of phononic excitations in a 1D Bose gas 
using the Truncated Wigner Approximation (TWA)~\cite{Steel-TWA}.
In this approach, the atomic field operators $\hat{\Psi}$ of Eq.~\eqref{eq:full_Hamiltonian} are replaced by classical variables $\Psi$, and products of these variables are identified with the corresponding fully symmetrised quantum operators.  A series of {\it ab initio} Monte Carlo simulations are performed.  Quantum indeterminacy appears through the statistical ensemble describing the initial state, whose probability distribution is identified with the (Gaussian) Wigner function.  The field is then evolved according to the dynamics of Hamiltonian~\eqref{eq:full_Hamiltonian}.  This is repeated for a large number of independent initial realisations, so as to get good statistics when computing averages.

\subsection{Reanalysing data of~\cite{Micheli-2022} with improved template}
\label{subsec:reanalysis}

Let us first reanalyse the series of simulations of parametric growth that were presented in~\cite{Micheli-2022}.  As mentioned above, in~\cite{Micheli-2022} the numerically observed occupation number $n_{k}(t)$ was fit to Eq.~\eqref{eq:old_n_template} in order to extract the effective dissipation rate.  This yielded results in which relatively strong deviations from prediction~(\ref{eq:Gamma_formula}) were observed (see Fig.~5 of~\cite{Micheli-2022}).

\subsubsection{Simulation parameters}
\label{sec:simulation_parameters}

We begin by listing the different parameters describing the behaviour of the gas.
First, the 1D gas' Hamiltonian dynamics is controlled by
\begin{itemize}
    \item its length $L$,
    \item the atoms' mass $m$,
    \item the strength of their interactions $g$,
\end{itemize}
while its quasi-condensed state is characterised by 
\begin{itemize}
    \item the background density $\rho_0$,
    \item the temperature $T$.
\end{itemize}
Second, the parameters controlling the modulation of the interaction in Eq.~\eqref{eq:g_modulation} are
\begin{itemize}
    \item the amplitude $a$,
    \item the frequency $\omega_p$,
    \item the duration $t_{\mathrm{max}}$.
\end{itemize}
Finally, the relevant parameters for our TWA simulations are 
\begin{itemize}
    \item the time step $\Delta t$,
    \item the spatial grid spacing $\Delta x$,
    \item the number of sites $n_x$,
    \item the number of realisations generated to compute 
    averages, $N_{r}$.
\end{itemize}

In the simulations of~\cite{Micheli-2022} the healing length $\xi = \hbar / m c$ (or equivalently the speed of sound $c$ since we will assume $m$ to be fixed) was fixed at initial time, and we thus use its initial value to adimensionalise the different parameters.
This series of simulation was generated by varying $\rho_0 \xi$~\footnote{Since $c^2 \propto g \rho_0$ is fixed, this means that $g$ is covaried with $\rho_0$. More directly, upon adimensionalisation $g m \xi / \hbar^2 = (\rho_0 \xi)^{-1}$.}. According to Eq.~(\ref{eq:Gamma_formula}), this induces a variation of the decay rate, which is inversely proportional to $\rho_{0}\xi$.  
The simulations are run for two distinct (pairs of) resonant modes: $k \xi = \pm 1.0$ and $k\xi = \pm 3.1$, which are selected by tuning 
the modulation frequency $\omega_{p} = 2\omega_{k}$; this requires $\omega_{p}t_{\xi} = 2.35$ and $\omega_{p}t_{\xi} = 11.15$, respectively.
Note that, for a fixed modulation amplitude $a$ of the interaction constant, the effective modulation amplitude $A_{k}$ of the mode frequency is $k$-dependent 
by virtue of Eq.~\eqref{eq:effective_amplitude_mod}, and the corresponding growth rates differ for $k\xi = 1.0$ and $k\xi = 3.1$. However, for each value of $k\xi$, the growth rate $G_k$ is kept constant so that, according to the free theory described by Eqs.~(\ref{eq:sol_nk_no_dissipation}) and~(\ref{eq:sol_ck_no_dissipation}), the behaviour of $n_k$ and $|c_k|$ should be identical. Any differences, therefore, must be due to interactions in the system\footnote{The parameters listed here fully characterise the 1D Bose gas. To make the link with the simulations presented in~\cite{Robertson-2018}, we note that a dimensional reduction from the 3D description of a gas in a cylindrical trap is performed there, so that the transverse trapping frequency $\omega_{\perp}$ and the associated harmonic oscillator width $a_{\perp} = \sqrt{\hbar/m\omega_{\perp}}$ (equal to the width of the gas only in the absence of atom-atom interactions) were explicitly mentioned.
In these simulations, the atomic field is adimensionalised by $\sqrt{\rho_0}$ so that $\tilde{g} = g\rho_{0}/\hbar\omega_{\perp}$ is the relevant adimensional coupling constant appearing in the code. There the speed of sound $c = \sqrt{\tilde{g}}$, but also the adimensional linear density $\rho_0 a_s$ (where $a_{s}$ is the $s$-wave scattering length, see footnote~\ref{fn:3d_to_1d}) were considered fixed. The key parameter used in~\cite{Robertson-2018} to govern the strength of phonon-phonon interactions was then $a_{s}/a_{\perp}$.
The relation $\rho_0 \xi = \rho_0 a_s (a_{s}/a_{\perp})^{-1} /\sqrt{\tilde{g}}$ shows that changing $a_{s}/a_{\perp}$ effectively amounts to changing the level of longitudinal interaction and dissipation, see Eq.~\eqref{eq:Gamma_formula}. }.  

All other parameters are fixed and we give their values in Tab.~\ref{tab:fixed_param}.

\begin{table}[!ht]
\begin{center}
\begin{NiceTabular}{|c|c|c|c|c|c|c|}
\Hline
 $ L / \xi$ & $k_{\mathrm{B}} T / m c^2$ & $a$  &  $\Delta t / t_{\xi}$ & $\Delta x / \xi$  & $n_x$ & $N_r$ \\ 
\Hline
\Hline
90.5 & 2 & 0.5  & 0.0225 & 0.35 & 256 & 400  \\ 
\Hline
\end{NiceTabular}
\vspace{5pt}
\caption{Parameters of the gas, modulation and simulation common to all TWA simulations of~\cite{Micheli-2022} reanalysed in the present work.
\label{tab:fixed_param}}
\end{center}
\end{table}

For the given length, the resonant modes at $k \xi =  1.0$ and $k\xi = 3.1$ correspond to the $i=15$ and $i=44$ discrete modes of the gas for $k = i 2 \pi / L $, respectively.
The corresponding growth rates are $G_{k}t_{\xi} = 0.06$ for $k\xi = 1.0$ and $G_{k}t_{\xi} = 0.10$ for $k\xi = 3.1$. The duration of modulation is $t/t_{\xi} = 22.5$ (respectively $t/t_{\xi} = 13.5$) for $k \xi = 1.0$ (resp. $k \xi = 3.1$), which when adimensionalised by the typical timescale given by the growth rate gives $t_{\rm max} G_{k} = 5.20$ (resp. $t_{\rm max} G_{k} = 5.66$).

\begin{table}[h!]
\begin{center}
\begin{NiceTabular}{|c|c|ccccccc|}[corners=NW]
\Hline
& $\rho_0 \xi$ & 399.3 & 199.7 & 133.1 &  99.8 & 66.6 &  49.9 &  33.3 \\ 
\Hline
\Hline
$\boldsymbol{k\xi = 1.0}$ & $\alpha_k = \Gamma_k/G_k$ & 0.068 & 0.136 & 0.204 & 0.2723 & 0.409 & 0.545 & 0.818 \\ 
& $\chi^2_{\nu}$ & 10.0 &  12.6 &  26.0 &  30.8 & 65.9 & 106.2 & 189.6 \\ 
\Hline
\Hline
$\boldsymbol{k\xi = 3.1}$ & $\alpha_k = \Gamma_k/G_k$ & 0.050 & 0.101 & 0.151 & 0.201 & 0.302 & 0.402 & 0.603  \\ 
& $\chi^2_{\nu}$ & 3.5 &  4.4 &  5.3 &  7.3 &  9.0 & 19.0 & 41.5 \\ 
\Hline
\end{NiceTabular}
\vspace{5pt}
\caption{Set of simulations performed in~\cite{Micheli-2022} and reanalysed in the present work. The 1D density parameter $\rho_{0}\xi$ is varied, while all other parameters are fixed to the values given in Sec.~\ref{sec:simulation_parameters}.  Runs are performed for two different values of the modulation frequency $\omega_{p}$, with resonant modes at $k\xi = 1.0$ and $k\xi = 3.1$.  Listed here are the predicted values of $\alpha_{k}$ (given prediction~(\ref{eq:Gamma_formula}) for $\Gamma_{k}$) and the reduced $\chi^2$ characterising the fit to template~(\ref{eq:n_c_initial_thermal}) which treats $\alpha$ as a fitting parameter. The best-fit values of $\Gamma$ are shown in Fig.~\ref{fig:fit_n_c_different_asap_early_times}, with details of the procedure in Sec.~\ref{subsec:fit}.
\label{tab:fig5_parameters}}
\end{center}
\end{table}

\begin{figure}
    \centering
\centering
\includegraphics[width=0.85\textwidth]{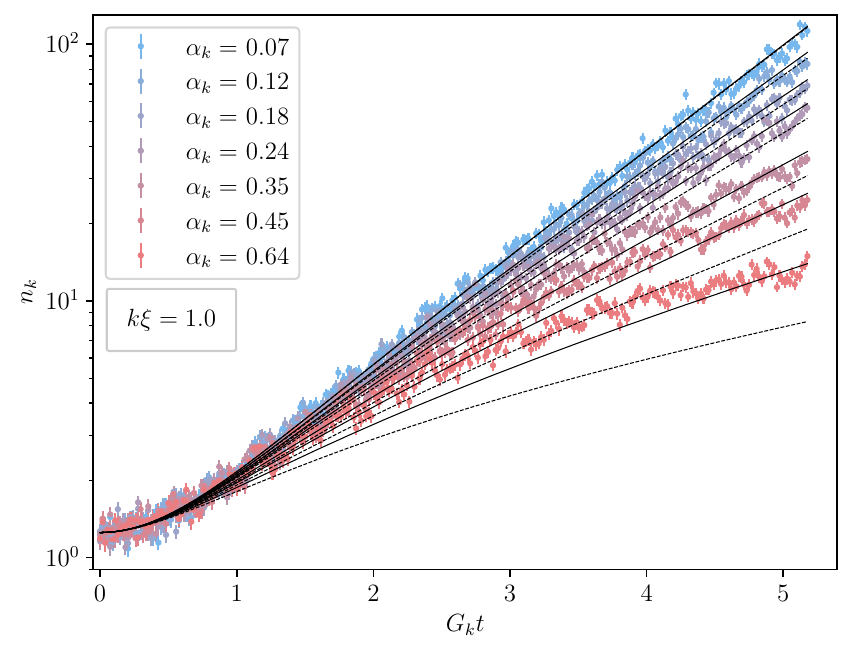}
\includegraphics[width=0.85\textwidth]{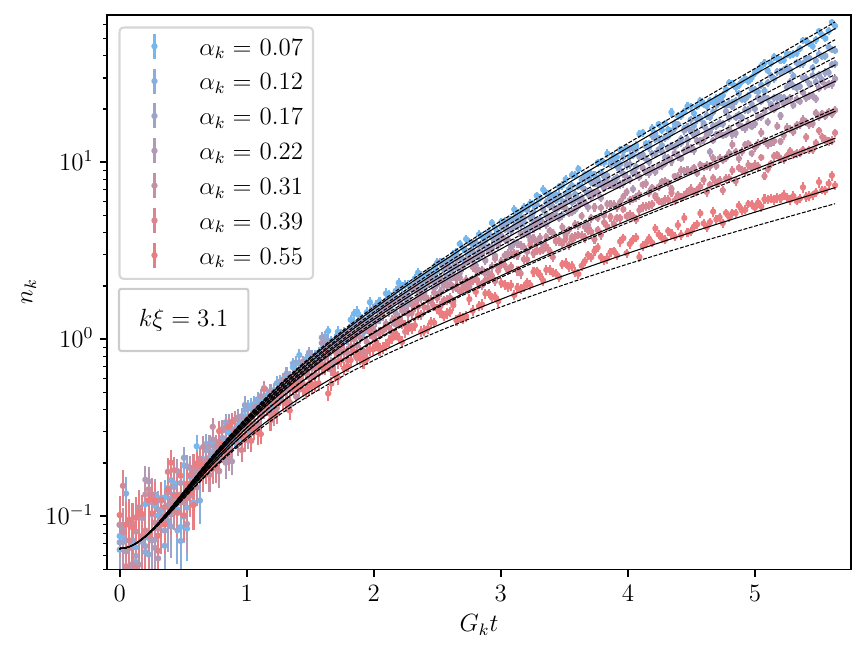}
\caption{Evolution of the number of excitations $n_k$ in the resonant mode with positive wavenumber $k \xi$ for different values of $\rho_{0} \xi$ as given in Tab.~\ref{tab:fig5_parameters}. The other parameters are fixed to the values given in Tab.~\ref{tab:fixed_param}. The colored dots with error bars correspond to the results of TWA simulations. The values of $\alpha_k$ reported are the best-fit values to template~(\ref{eq:n_c_initial_thermal}) with details of the procedure in Sec.~\ref{subsec:fit}. The full (resp. dashed) curves represent the evolution predicted by template~(\ref{eq:n_c_initial_thermal}) for best-fit values (resp. predicted values given in Tab.~\ref{tab:fig5_parameters}) of $\alpha_k$.}
\label{fig:nk_different_asap}
\end{figure}

\subsubsection{Behaviour of $n_{k}$ and $\Delta_{k}$}
\label{subsec:n_delta}

Numerical results for the evolution of the system in this series of simulations are shown in Figs.~\ref{fig:nk_different_asap} where we plot the occupation $n_{k}$ of the resonant mode (with positive wavenumber) as a function of $\tau = G_{k}t$, the same adimensionalised time used in Eqs.~(\ref{eq:n_res}) and~(\ref{eq:c_res}).
The upper and lower panels correspond to $k\xi = 1.0$ and $k\xi = 3.1$, respectively.  The data points are numerically extracted by averaging over $N_{r} = 400$ realisations, while the associated errors are given by $\sqrt{{\rm Var}(n_{k})/N_{r}}$ where ${\rm Var}(n_{k})$ is the variance of $n_{k}$.  (More details on the error analysis of the simulations can be found in the appendices.)  The dashed curves show prediction~(\ref{eq:n_res}), given the corresponding values of $\alpha_{k}$ in Tab.~\ref{tab:fig5_parameters}.  
Generally speaking, the numerical observations are in good agreement with the theoretical prediction.  We indeed see that, with decreasing $\rho_{0}\xi$, the parametric growth is slowed down by an amount consistent with the corresponding prediction for $\alpha_k$.  We see no sign of saturation, consistent with all predicted values of $\alpha_k$ being smaller than $1$.
That being said, we do notice some degree of discrepancy. As had already been noted in~\cite{Micheli-2022}, the discrepancies are more noticeable for $k\xi = 1.0$, with our prediction seeming to overestimate the effects of dissipation.

\begin{figure}
    \centering
\centering
\includegraphics[width=0.75\textwidth]{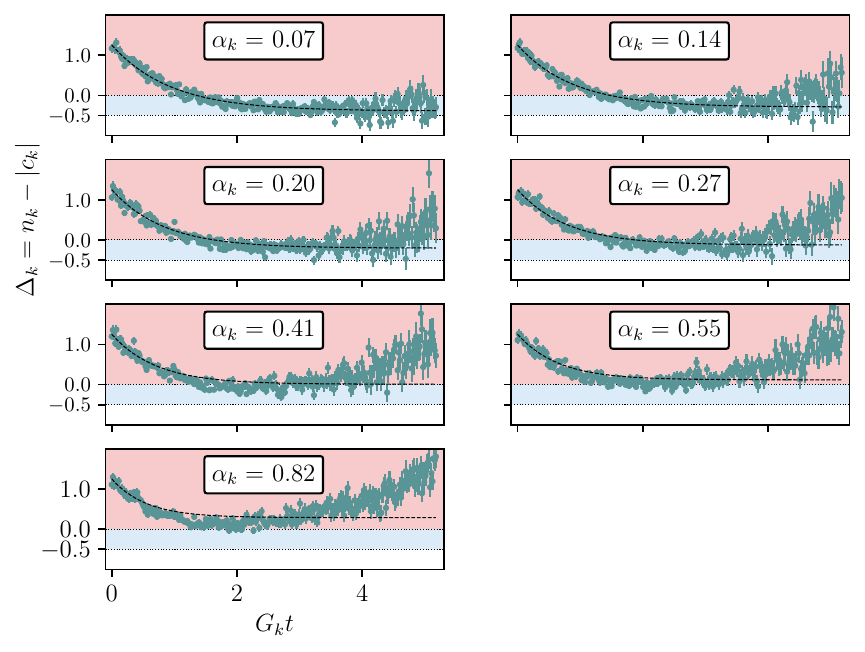}
\includegraphics[width=0.75\textwidth]{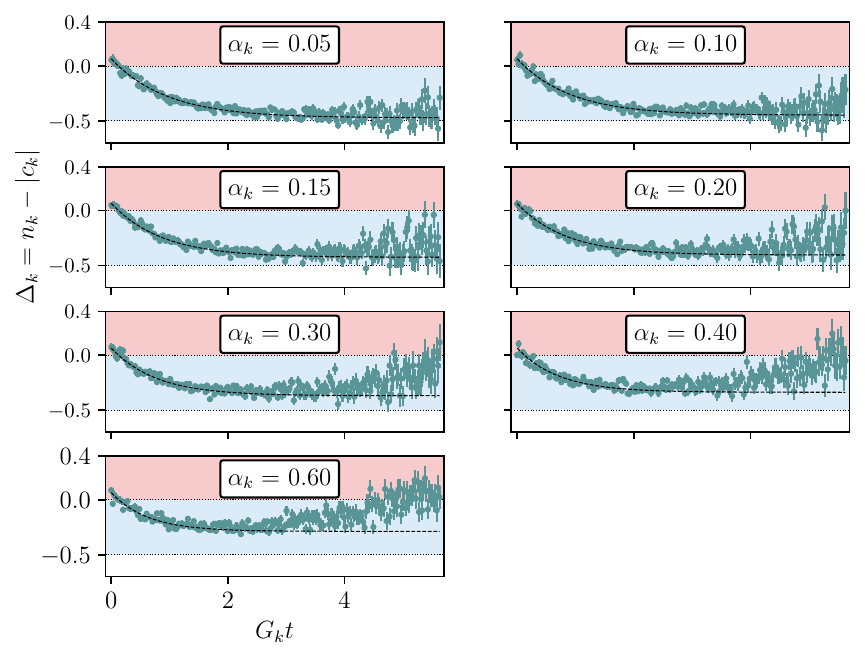}
\caption{Evolution of the nonseparability parameter $\Delta_k$ of the resonant modes during modulation as a function of adimensionalised time $G_k t$. Green dots show results of TWA simulations for different gas density $\rho_0 \xi$, so different expected $\alpha_k$ computed from Eq.~\eqref{eq:Gamma_formula}. Upper and lower panels correspond respectively to resonant modes $k \xi = \pm 1.0$ and $k \xi = \pm 3.1$.
Red and blue shaded regions correspond to $\Delta_k > 0$, separable states, and $-0.5 < \Delta_k < 0$, entangled states. The region $\Delta_k < -0.5$ is left blank as it should be excluded for physical states. Finite statistics might still lead to points in the region.
Relevant parameters are listed in the figure and in Tabs.~\ref{tab:fixed_param}-\ref{tab:fig5_parameters}.
The error bars correspond to one standard deviation on each side of the mean value, see App.~\ref{app:error} for more details. Dashed lines show predictions~\eqref{eq:n_c_initial_thermal} using the value of $\alpha_k$ quoted in the figure.}
\label{fig:Delta_different_asap}
\end{figure}

While~\cite{Micheli-2022} dealt only with the occupation number $n_k$, we now also want to discuss the degree of correlation between the excitations of opposite momentum $c_k$.
To this end, we plot the behaviour of $\Delta_k = n_k - |c_k|$, which is both physically meaningful since $\Delta < 0$ demonstrates nonseparability, and typically small so that deviations from predictions should be easily and clearly observed. Note that in our simulations we compute $\Delta_k$ using $n_k$, the occupation number of the resonant mode with \textit{positive} wavenumber; since by statistical isotropy we expect $n_{k} = n_{-k}$ for a large number of realisations, this is a matter of convention.
The results for $\Delta_{k}(t)$ are shown in Fig.~\ref{fig:Delta_different_asap}, for the same set of simulations as Fig.~\ref{fig:nk_different_asap}.  
Data points are again shown for the numerically extracted values, with their corresponding error bars.  
(The error analysis is a bit more involved for $\Delta_{k}$; see the appendices for details.)
Dashed curves show the analytical prediction given by the second of Eqs.~(\ref{eq:sol_nk_ck_exact_resonance}).
Generally speaking, the prediction agrees very well with observations, and in particular the predicted drift of $\Delta_{k}$ away from $-1/2$ as $\alpha_{k}$ is increased is corroborated by the numerical observations (this is especially clear for $k\xi = 3.1$).
The main discrepancies are twofold.
First, as for $n_{k}$, they are more noticeable for $k\xi = 1.0$, where the effects of dissipation tend to be overestimated by the prediction: the numerically observed $\Delta_{k}$ actually dips a little below what is predicted.
Second, whereas our model predicts a monotonic decrease of $\Delta_k$ to the asymptotic value $\Delta_{\infty}$ of Eq.~(\ref{eq:asymptotic_Delta}), the simulations indicate that at sufficiently late times $\Delta_k$ grows again, even becoming positive so that the nonseparability is lost.
This behaviour was already noted in~\cite{Robertson-2018}, and points to the existence of an additional decoherence mechanism that is not captured by our current model\footnote{In~\cite{Robertson-2018} this was suggested as the first clear signal of phonon-phonon interactions, but our current results compel us to revise this, since the reduction of the growth rate and the saturation of $\Delta$ at a value above $-1/2$ due to dissipative processes already occur before $\Delta$ has the chance to increase and become positive.}.
For all the simulations shown here, the predicted values of $\alpha_k$ are smaller than 1, so we are never in the regime of saturated growth.  However, for $k\xi = 1.0$, we {\it do} cross the threshold for nonseparability of the final state, since the thermal population $n_{\rm th} = 1.25$ is quite significant and the nonseparability threshold occurs at $\alpha_k = 1/2n_{\rm th} = 0.40$.
Indeed, despite our model overestimating dissipation at early times for $k\xi = 1.0$, we observe that for the three values of $\alpha_k$ which lie above $0.40$ in the top panel of Fig.~\ref{fig:Delta_different_asap}, nonseparability is at best only barely reached. 

\subsubsection{Extracting best-fit values of the decay rate}
\label{subsec:fit}

\begin{figure}
    \centering
\centering
\includegraphics[width=0.9\textwidth]{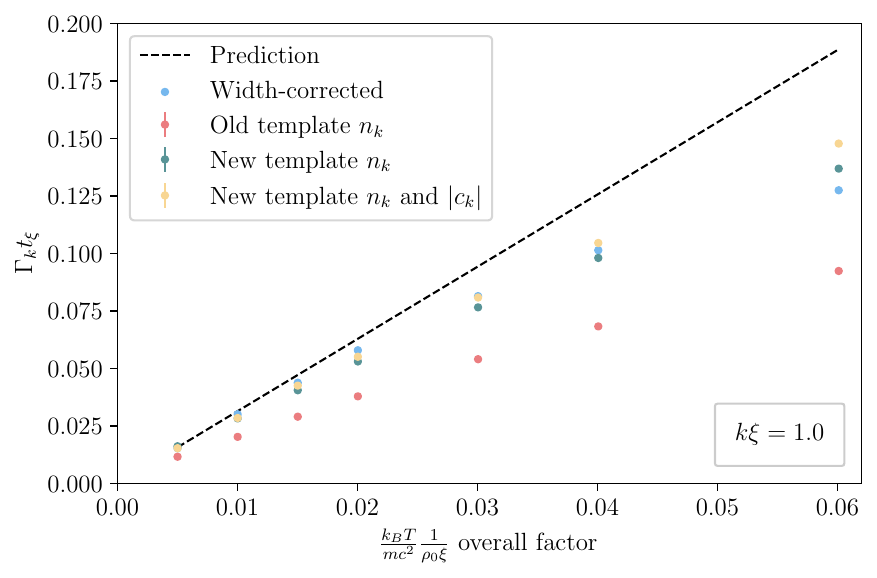}
\includegraphics[width=0.9\textwidth]{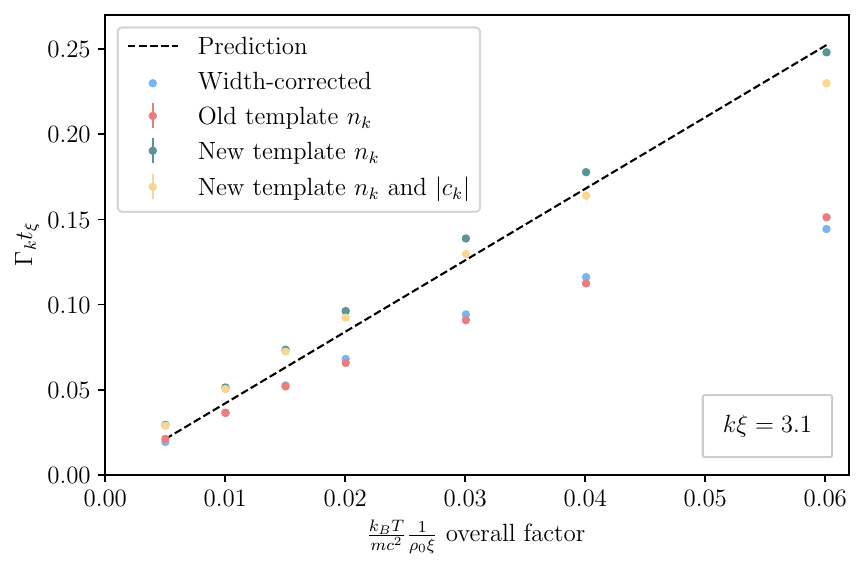}
\caption{ Best-fit values for the quasiparticle dissipation rate $\Gamma_k$ extracted from TWA simulations data for two different pairs of resonant modes $k \xi = \pm 1.0$ for the upper panel and $k \xi = \pm 3.1$ for the lower panel. These values are compared to the prediction~\eqref{eq:Gamma_formula}. The figure is an updated version of Fig.~5 in~\cite{Micheli-2022}.
The value of the reduced $\chi^2$ for the fits performed jointly over $n_k$ (for the mode with positive wavenumber) and $|c_k|$ are reported in Tab.~\ref{tab:fig5_parameters}.
\label{fig:fit_n_c_different_asap_early_times}
}
\end{figure}

To complete the reanalysis, we wish to test how much the use of template~(\ref{eq:old_n_template}) to fit the evolution of $n_k$ in~\cite{Micheli-2022} skewed our estimate of $\Gamma_k$.
To that end we redo the fitting analysis, this time using the refined analytical prediction~(\ref{eq:sol_nk_ck_exact_resonance}).
Given that in the present work we consider both the evolution of $n_{k}$ and $\left|c_{k}\right|$, we perform the fit in two ways: using $n_{k}$ alone (as in~\cite{Micheli-2022}), and using both $n_{k}$ and $\Delta_{k}$ simultaneously to make a joint fit\footnote{These joint fits are performed by minimising the sum of the squared residuals weighted by the covariance matrix of the errors in $n_k$ and $|c_k|$. The procedure is detailed in App.~\ref{app:error}. The resulting optimal values are comparable with those obtained using the routine {\bf curve\_fit} of the Python package {\bf scipy}.}. 
The best-fit value of $\alpha_k$ is then converted into a value for the dissipation rate $\Gamma_k$.
The results are shown in Fig.~\ref{fig:fit_n_c_different_asap_early_times}, which is an updated version of Fig.~5 of~\cite{Micheli-2022} that now features the results of the fit with our new template. We see that the new template does provide an improvement in that the disagreement between the best-fit values and our refined predictions is reduced.
Moreover, the fits obtained using either $n_{k}$ alone or $n_{k}$ and $\left|c_{k}\right|$ simultaneously are in good agreement, indicating consistency of our expressions for $n_{k}$ and $\left|c_{k}\right|$ and showing that it is really the refined template that is responsible for the better agreement.
In particular, for $k \xi = 3.1$ the disagreement is now very small across the whole range of interactions. 
On the other hand, for $k \xi = 1.0$ we again find discrepancies.
The inadequacy of the template is also manifest in the values of the reduced $\chi^2$ reported in Fig.~\ref{fig:fit_n_c_different_asap_early_times}, which is on the order of unity for $k \xi = 3.1$ and significantly larger for $k \xi = 1.0$.  
This suggests that the visble discrepancies between theory and observation seen for $k\xi = 1.0$ are not simply due to an inaccurate prediction for $\Gamma_{k}$, but that the mechanisms at play behave in a qualitatively differently manner than in our model, and that a different template would be more appropriate.

\subsubsection{Discussion on deviations from the prediction}

A noticeable deviation from our predictions is the non-monotonicity of $\Delta_k$, with nonseparability being eventually lost at late times for large enough interactions. This behaviour indicates that some additional dissipative processes are at play, {\it e.g.}, a degree of nonlinear damping once the number of produced phonons becomes sufficiently large.
On the other hand, we do not have a conclusive answer as to what causes the greater discrepancies seen for $k\xi=1.0$ as compared to $k\xi=3.1$.  However, there were two deviations noted and discussed in~\cite{Micheli-2022} that should be mentioned here, as they may hint at possible explanations.

First, it was numerically observed in~\cite{Micheli-2022} that the population of a single mode which is instantaneously excited (rather than being excited by parametric resonance) would decay exponentially only after a certain ``critical time''  $t_{\mathrm{crit}}$ has passed.  During an initial phase, the decay proceeds quadratically in time, and is much slower than would be suggested by the exponential decay rate~(\ref{eq:Gamma_formula}).  This critical time varies roughly as $t_{\mathrm{crit}} \propto 1/\left(k\xi\right)^{3}$, and is therefore much more significant for $k\xi=1.0$ than for $k\xi=3.1$.  This could help to explain the larger discrepancies seen in our simulations for $k\xi=1.0$.  
However, while this early-time deviation has been observed and understood in the case where the condensate is otherwise stationary,  
it is less clear how or whether this critical time is relevant in the case of parametric resonance where phonons are being continuously injected into the mode in question.

Second, and as already mentioned, significant deviations in the extracted value of $\Gamma_{k}$ had already been observed in~\cite{Micheli-2022}, and in the absence of the refined template~(\ref{eq:n_res}) another explanation for the deviations was proposed.  It was noted that, when one considers a resonant peak with a finite width, then nearby (slightly off-resonant) modes can feed back into the central mode and thereby act against the dissipation, effectively reducing the rate $\Gamma_k$.  A self-consistent description of this effect was derived in Appendix~B of~\cite{Micheli-2022}, turning the observed width of the resonant peak into a correction for the decay rate.  The width-corrected predictions for $\Gamma_{k}$ are reproduced in Fig.~\ref{fig:fit_n_c_different_asap_early_times}.  Remarkably, these came very close to the old extractions of $\Gamma_{k}$ (based on template~(\ref{eq:old_n_template})) for $k\xi=3.1$, boosting the plausibility of this explanation.  For $k\xi=1.0$, there remained a significant discrepancy even with the width-corrected predictions, but there are reasons for which that could be true (such as the critical time issue mentioned above).  However -- and somewhat intriguingly -- the refined analysis performed here muddies the waters as far as the effect of the peak width is concerned.  For while the newly extracted decay rates come very close to the width-corrected predictions for $k\xi=1.0$ -- suggesting that this effect is indeed slowing the decay in the expected way -- the new analysis of $k\xi=3.1$ yields decay rates very close to the predicted values when the finite-width correction is {\it not} taken into account.  So there are some conflicting observations concerning these corrections, and it is not yet clear how they are to be reconciled.

\subsection{Demonstrating the two growth regimes}
\label{subsec:2_growth}

The second goal of our numerical studies is to 
extend the range of $\alpha$ probed by simulations as compared to~\cite{Micheli-2022}. In particular we want to explore the two regimes of extended exponential growth and saturated growth, predicted to occur for $\alpha < 1$ and $\alpha > 1$, respectively\footnote{Note that the extended exponential growth will also lead eventually to saturation, as was numerically observed in~\cite{Robertson-2018}.  This late-time saturation is driven by nonlinearities due to a large occupation of the resonant modes, rather than by the linear dissipation rate applicable at small occupation numbers.}.  

For this series of simulations we use parameters that are currently achievable in experiment\footnote{The values quoted are derived from the following experimental parameters. The temperature $T = 90\,{\rm nK}$.  The transverse trapping frequency is $\omega_{\perp}/2\pi = 1650\,{\rm Hz}$, which for metastable helium yields $a_{s}/a_{\perp} \approx 0.006$.  The speed of sound achievable at the centre of the gas is $c = 16.5\,{\rm cm}/{\rm s}$, which gives $\tilde{g} = mc^{2}/\hbar\omega_{\perp} \approx 1.65$. Using an approximation of $G(\rho_0 a_s)$ given in~\cite{Robertson-2018}, we can infer from the value of $\tilde{g}$ that of the density of the gas $\rho_0 a_s$. For $\tilde{g}= 1.65$, we get $\rho_0 a_s = 1.97$, and so $\rho_0 \xi \approx 255.2$. Note that the inversion formula from $\tilde{g}$ to $\rho_0 a_s$ is quite sensitive to small differences, so that not much attention should be payed to the exact values for these parameters but rather to their variation.}. The values of all fixed parameters are summarised in Tab.~\ref{tab:omegamod3_fixed_parameters}.

\begin{table}[h!]
\begin{center}
\begin{NiceTabular}{|c|c|c|c|c|c|c|c|}
\Hline
 $ L / \xi$ & $k_{\mathrm{B}} T / m c^2$ & $\rho_0 \xi $ & $\omega_{p} t_{\xi} $ & $\Delta x / \xi$ &  $\Delta t / t_{\xi}$ & $n_x$ & $N_r$ \\ 
\Hline
\Hline
329 & 0.69 & 255.2 & 4.97 & 0.64 & 0.0064 & 512 & 400  \\ 
\Hline
\end{NiceTabular}
\vspace{5pt}
\caption{
Parameters used for the new series of simulations exploring a large of $\alpha_k$ by varying the growth rate $G_k t_{\xi}$ with values given in Tab.~\ref{tab:omegamod3_varyingA_parameters}.
\label{tab:omegamod3_fixed_parameters} }
\end{center}
\end{table}

The quoted value of the modulation frequency $\omega_p$ was tuned to be exactly twice that of a mode ($\omega_p = 2 \omega_k$) to avoid any deviations from our prediction due to the amplified modes being slightly off-resonance. We choose the resonant modes at $k \xi = \pm 0.84$ so that the thermal population there is not too large: $n_{k}^{\mathrm{th}} = 0.36$. This allows the achievement of entanglement for a large range of dissipation: $\alpha_k < 1 / 2 n_{k}^{\mathrm{th}} \approx 1.37$. Note in particular that, since $n_{k}^{\rm th} < 1/2$, it allows for achievement of nonseparability even when $n_{k}$ saturates.
Finally, to make manifest the effect of dissipation we are going to co-vary the two remaining parameters: the excitation amplitude $a$ and the duration of modulation $t_{\mathrm{max}}/t_{\xi}$ while keeping $a t_{\mathrm{max}}$, and so $G_k t_{\mathrm{max}}$, fixed.
Since the free theory Eqs.~(\ref{eq:sol_nk_no_dissipation})-(\ref{eq:sol_ck_no_dissipation}) only depends on the combination $G_k t$, all curves plotted as a function of time adimensionalised by their respective $G_k$ should be on top of each other when dissipation is absent. Again, any difference is due to the presence of interactions.
We pick $G_k t_{\mathrm{max}} = 2.55$ so that in the free theory $n_{\mathrm{max}} \approx 5.1$, and we thus remain in the regime of small occupation number and thus avoid any nonlinearities in $n_{k}$. 

\begin{table}[h!]
\begin{center}
\begin{tabular}{|c|| c| c |c| c| c| c |} 
 \hline
 $a$  & 0.239 & 0.106 & 0.0636 & 0.0203 & 0.0136 & 0.0106 \\ [0.5ex] 
 \hline
 $t_{\mathrm{max}} / t_{\xi}$ & 27.6 &  62.0 & 103 & 324 &  482 & 620 \\ [0.5ex] 
 \hline 
$\alpha_k$  & 0.09 & 0.20 & 0.34 & 1.06 & 1.59 & 2.03 \\
 \hline 
\end{tabular}
\vspace{5pt}
\caption{Values taken by the parameters which are varied in this series of simulations.
\label{tab:omegamod3_varyingA_parameters}}
\end{center}
\end{table}

\begin{figure}
    \centering
\centering
\includegraphics[width=0.95\textwidth]{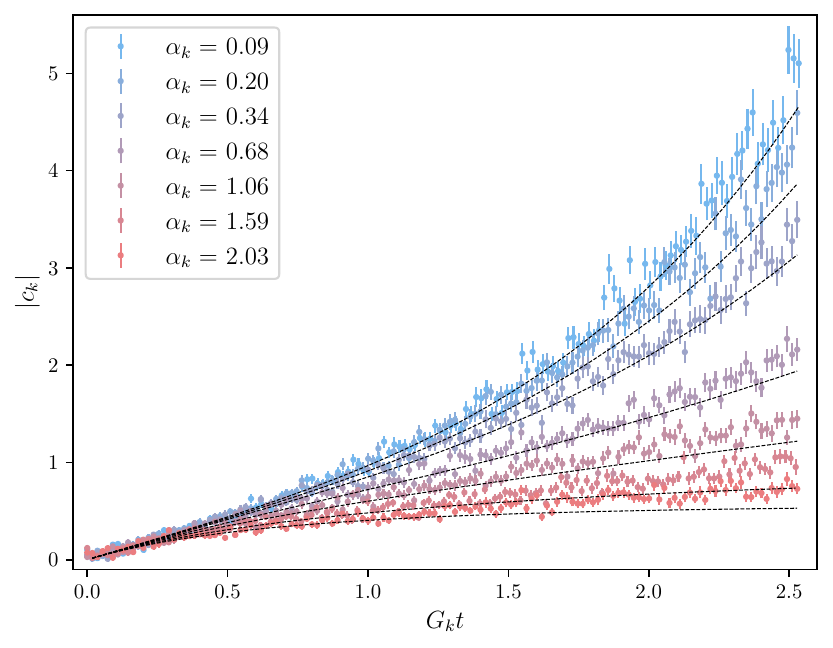}
\caption{Evolution of the pair correlation $|c_k|$ of the resonant modes during modulation as a function of adimensionalised time $G_k t$. Coloured dots show the results of TWA simulations for different modulation amplitudes $a$, so different expected $\alpha_k$ computed from Eq.~\eqref{eq:Gamma_formula}. The amplitude decreases from blue to red. Relevant parameters are listed in Tabs.~\ref{tab:omegamod3_fixed_parameters}-\ref{tab:omegamod3_varyingA_parameters}. The error bars correspond to one standard deviation on each side of the mean value, see App.~\ref{app:error} for more details. Dashed lines are predictions~\eqref{eq:n_c_initial_thermal} using the value of $\alpha_k$ quoted in the figure. }
\label{fig:ck_different_A}
\end{figure}

\begin{figure}
    \centering
\centering
\includegraphics[width=0.95\textwidth]{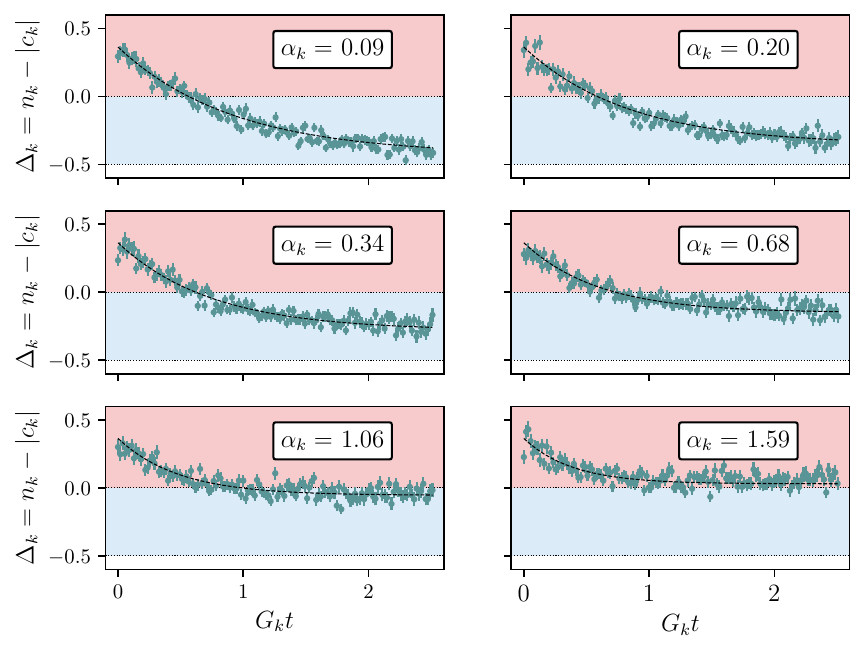}
\caption{Evolution of the nonseparability parameter $\Delta_k$ of the resonant modes during modulation as a function of adimensionalised time $G_k t$. Green points correspond to results of TWA simulations for different modulation amplitudes $a$, so different expected $\alpha_k$ computed from Eq.~\eqref{eq:Gamma_formula}. 
Red and blue shaded regions correspond to $\Delta_k > 0$ (separable states) and $-0.5 < \Delta_k < 0$ (entangled states). The region $\Delta_k < -0.5$ is left blank as it should be excluded for physical states, though finite statistics might still lead to points in the region.
Relevant parameters are listed in the figure or in the text of Sec.~\ref{subsec:2_growth}. The error bars correspond to one standard deviation on each side of the mean value, see App.~\ref{app:error} for more details. Dashed lines are predictions of Eq.~\eqref{eq:n_c_initial_thermal} using the value of $\alpha_k$ quoted in the figure. }
\label{fig:Delta_different_A}
\end{figure}

Figures~\ref{fig:ck_different_A} and~\ref{fig:Delta_different_A} show the behaviour of $|c_k|$ and $\Delta_k = n_k - |c_k|$ as a function of time. The behaviour of $n_k$ is close enough to that of $|c_k|$ not to give any additional information so we do not show it.
A few observations are in order.
First, Fig.~\ref{fig:ck_different_A} shows that, even for growth rates smaller than the dissipation rate ($\alpha_k > 1$), the correlation amplitude $|c_k|$ (and the occupation number $n_k$) still grows, corroborating our analytical predictions.  (The origin of this growth in source terms, both quantum and classical, was discussed in Sec.~\ref{subsubsec:growth_and_saturation}.) 
Second, the three curves corresponding to values $\alpha_k > 1$ clearly do not grow exponentially as their slopes are decreasing, in contrast to those for $\alpha_k <1$. We have thus numerically demonstrated the existence of two distinct regimes of growth, as predicted in~\cite{Busch-2014} and reviewed above. 
In addition, the saturated regime was attained here while considering experimentally motivated values of the temperature and of the gas' characteristics, so this regime might actually be experimentally achieved.
Finally, Fig.~\ref{fig:Delta_different_A} shows, in agreement with our model and in a clearer way than the data shown in Sec.~\ref{subsec:reanalysis}, that for $\alpha_k >  1/ 2 n^{\mathrm{th}}_{k_\mathrm{eq}} \approx 1.37$ we never reach an entangled state in the TWA simulations.

Overall, for the set of parameters given in Tab.~\ref{tab:omegamod3_varyingA_parameters}, the agreement between our predictions and the results of TWA simulations is very good, and better than for the set of simulations of~\cite{Micheli-2022} reanalysed in the previous section. 
In particular, in Fig.~\ref{fig:Delta_different_A} we do not witness any clear early or late-time deviations in the behaviour of $\Delta_k$. 
This might be due to the use of a shorter simulations in adimensionalised time $G_k t_{\mathrm{max}} = 2.55$, while in the data of~\cite{Micheli-2022} late-time deviations only appeared for $G_k t_{\mathrm{max}} \geq 2$.


\section{Summary and conclusion
\label{sec:Conclusion}}

In this article, we have reviewed some previous theoretical results concerning parametric amplification of phonons in a modulated one-dimensional Bose gas.  In particular, a phenomenological description of the evolution in the presence of weak dissipation was given in~\cite{Busch-2014}, though the specific phenomena described had not previously been tested in fully nonlinear simulations.  In addition, a microphysical description of phonon decay through interaction with a thermal bath was given in~\cite{Micheli-2022}, and while some comparisons with fully nonlinear simulations were made, the application to the phenomena described in~\cite{Busch-2014} was not.  Here, we have closed the circle: we have shown that fully nonlinear simulations of the Bose gas do indeed show much of the phenomenology predicted in~\cite{Busch-2014}, and that the strength of the relevant dissipation acting on the number of produced quasiparticles $n_k$ and their correlation $|c_k|$ is indeed well-described by the prediction of~\cite{Micheli-2022}.

Yet, this study highlights some limitations of the phenomenological framework presented in~\cite{Busch-2014}, as well as the limited validity of the dissipation rate computed in~\cite{Micheli-2022}. First, \cite{Busch-2014} predicts that the nonseparability parameter $\Delta_k = n_k-\left|c_k\right|$ decreases monotonically in time, and therefore that any nonseparability achieved by the evolution endures as long as the modulation continues.  As had already been noticed in~\cite{Robertson-2018}, the fully nonlinear simulations show that this behaviour does not persist indefinitely, with $\Delta_k$ eventually increasing to become positive. 
This indicates an additional source of decoherence at play which would slow the growth of $|c_k|$ with respect to that of $n_k$. 
Second, we also see some discrepancies between our model and the results of numerical simulations at early times for the series of values of~\cite{Micheli-2022} with resonant modes $k \xi = \pm 1.0$
These were already noticed in~\cite{Micheli-2022} and were blamed on the existence of a critical time, scaling as $1/(k \xi)^3$, below which our description was not valid. However, in Sec.~\ref{subsec:2_growth} we show results of simulations where the resonant modes are located at $k \xi = \pm 0.84$ which do not exhibit early-time deviations. This shows that the regime of validity in $k \xi$ of the exponential dissipation we predict might have to be reviewed in the presence of an oscillating background, or might also depend on other parameters such as the temperature and the geometry of the gas.


\section*{Acknowledgments}

In the spirit of this special issue of the {\it Comptes Rendus}, we dedicate this article to our friend and mentor Renaud Parentani, who initiated and led the theoretical endeavour for the analysis of the experiment in~\cite{Jaskula2012}.  This work is a natural continuation of his ideas, and its authors owe much to his flair for precise thinking and enthusiastic discussions.  His last months of working with us were directed towards a determination of the phonon damping rate in a 1D quasicondensate, and we believe he would be pleased with the results presented here.
We also thank the experimental team of the COSQUA project (Rui Dias, Clothilde Lamirault, Charlie Leprince, Quentin Marolleau, Denis Boiron, Chris Westbrook, and especially Victor Gondret) for constant feedbacks and insightful discussions.
A. M. thanks Catherine Beauchemin and Enrico Rinaldi for explaining some MCMC basics, and Catherine Beauchemin for discussing at length statistical checks. A. M. also wants to thank all iTHEMS members, in particular Tsukasa Tada, for providing a supportive research environment.


\section{Appendix : Error estimation}

In this appendix, we detail and check part of the assumptions made to derive the error bars presented in the plots of the main text. We first discuss the error estimation of the number of quasiparticles in the resonant modes, $n_{\pm k}$, and the correlation amplitude, $ c_k = c_{k, R} + i c_{k, I}$, which are expected to be normally distributed due to the Central Limit Theorem (CLT)~\cite{ashProbabilityMeasureTheory2007}. Building on this, we then move on to the error estimation of the magnitude of the correlation $\lvert c_k \rvert $ and the difference $\Delta_k = n_k - \lvert c_k \rvert$. These are extensively used in the paper, but as we detail below, the errors on these quantities have, \textit{a priori}, no reason to be normally distributed. Then, in the third section we detail the fitting procedure used in Sec.~\ref{subsec:fit} to extract the value of the decay rate $\Gamma_k$ and discuss the meaning of the error bars.  Finally, in the fourth section (which can be ignored in a first reading), we perform further \textit{joint} normality checks for the estimators.

\subsection{Errors on basic quantities satisfying CLT}
\label{app:error}

\subsubsection{Estimators for basic quantities}

We are modelling a quantum system for which predicted quantities correspond to \textit{ensemble averages}. In particular, our model~(\ref{eq:eomncdissipapprox}) describes the evolution of the average number of quasiparticles in the resonant modes $n_{k}$ and their average correlation amplitude $c_k$ -- {\it not} the values of these quantities in any particular realisation of the experiment.
We estimate these average values in our TWA simulations by sampling $N_{r}$ random initial conditions, corresponding approximately to the initial thermal state of the gas, and evolving these realisations using the nonlinear classical equation of motion for the atomic field $\Psi$ given by the Hamiltonian~\eqref{eq:full_Hamiltonian} which corresponds to the Gross-Pitaevskii equation. The values of the atomic field $\Psi^{(i)} (x,t)$ for each realisation, at each point of the grid and each time, are saved. The relevant average values are estimated by averaging over the values obtained in each realisation, {\it e.g.}
\begin{equation}
\label{def:estimator_avg_n}
    \left \langle n_k \right \rangle  \approx \bar{n}_k  = \frac{1}{N_{r}} \sum_{i=1}^{N_{r}} b^{(i)  \star}_k b^{(i)}_k - \frac{1}{2} \, ,
\end{equation}
where $\left\langle n_{k}\right\rangle$ is the true ensemble average while $\bar{n}_{k}$ is the estimator found by averaging over a finite number of numerical realisations\footnote{Note that the subtraction of $1/2$ comes from the identification of TWA averages with those of {\it symmetrized} quantum operators, and since $\left(\hat{b}_{k}\hat{b}_{k}^{\dagger}+\hat{b}_{k}^{\dagger}\hat{b}_{k}\right)/2 = \hat{n}_{k} + 1/2$ this entails that we must subtract $1/2$ to get $\left\langle\hat{n}_{k}\right\rangle$.}.
How do we then estimate the typical error $\delta \langle n_k \rangle$ made when approximating $\langle n_k \rangle$ by $\bar{n}_k$?
First, note that Eq.~\eqref{def:estimator_avg_n} defines our estimator $\bar{n}_k$ as the (normalised) sum of independent and identically distributed (i.i.d.) random variables $n^{(i)}_k = b^{(i)  \star}_k b^{(i)}_k - 0.5$ with expectation value $\langle n_k \rangle$ and variance $\mathrm{Var} \left[ n_k \right]$. The CLT then asserts that $\sqrt{N_{r}} (\bar{n}_k - \langle n_k \rangle)$ approaches a centred normal distribution with variance $\mathrm{Var} [ n_k ] = \langle n_k^2 \rangle - \langle n_k \rangle ^2 = \sigma^{2}(n_k)$, {\it i.e.}, the estimator is asymptotically normal.
Assuming $N_{r}$ is large enough, $\bar{n}_k$ will thus be normally distributed with mean $\langle n_k \rangle$ and variance $\mathrm{Var} \left[ n_k \right]/N_{r}$. A good estimation of the typical error made when approximating $\langle n_k \rangle$ by $\bar{n}_k$ is thus given by $\sigma(n_k) / \sqrt{N_{r}}$. We are therefore left with estimating $\sigma(n_k)$, the standard deviation of $n_k$. This can be done using the standard unbiased estimator of the variance of a random variable $x$ from a sample
\begin{equation}
\label{def:estimator_variance}
S_{x} = \frac{1}{N_{r} -1} \sum_{i=1}^{N_{r}} \left( x^{(i)} - \bar{x} \right)^2 \, .
\end{equation}
We thus take the typical error made on $\bar{n}_k$ to be $\delta \langle n_k \rangle = \sqrt{S_{n_k} / N_{r}}$. We have used this estimate for the error bars 
in the figures of the main text showing the occupation number, {\it e.g.}, Fig.~\ref{fig:nk_different_asap}.

Note that the CLT applies to 
any i.i.d. random variables, irrespective of their distribution. In particular, it applies to $n^{(i)}_k$ for any of the modes $k$, at any time, and to 
\begin{align}
\label{def:estimator_avg_re_im_c}
\bar{c}_{k,R} = \frac{1}{N_{r}} \sum_{i=1}^{N_{r}} \mathrm{Re} \left[ b^{(i)}_k b^{(i)}_{-k} \right] \, , \qquad
\bar{c}_{k,I} = \frac{1}{N_{r}} \sum_{i=1}^{N_{r}} \mathrm{Im} \left[  b^{(i)}_k b^{(i)}_{-k} \right] \, ,
\end{align}
estimators of $c_{k,R}$ and $c_{k,I}$, the real and imaginary parts of $c_k = c_{k,R} + i c_{k,I}$. The variances of these quantities are also estimated using Eq.~\eqref{def:estimator_variance}.

\subsubsection{Normality checks for basic quantities}
\label{app:normal_basic}

To ensure that our estimated uncertainties are meaningful, we should check that the number of realisations $N_{r}$ is sufficiently large for the probability distribution of $\bar{n}_{\pm k}$ to be well-approximated by a normal distribution, with the expected mean $\langle n_{\pm k} \rangle$ and variance $\mathrm{Var} [ n_{\pm k} ]$. 
A direct but costly approach would be to redraw $N_{r}$ independent realisations and compute $\bar{n}_{\pm k}$ a large number of times $N_s$, and plot the resulting distribution.
For large $N_{r}$ this is impractical since it would require the generation and evolution of $N_s \times N_{r}$ realisations. Instead, we perform checks using only the $N_{r}$ realisations of $n_{\pm k}^{(i)}$ that we already have at our disposal. The idea is that, if $N_{r}$ is large enough, our sample should be a good representation of the underlying probability distribution function (PDF). Following the {\it bootstrap} approach~\cite{EfronBootstrap}, we draw random samples of $l < N_{r}$ realisations from this set as a proxy for performing many runs of $l$ simulations.
We allow ourselves to include the same realisation several times within a sample, 
for otherwise the possible subsamples would become very restricted at $l \sim N_{r}$ and their statistics would be biased towards those of the specific full sample, away from the underlying probability distribution. 
From the subsamples thus obtained we compute the estimators $\bar{n}_{\pm k} (l)$ and $S_{n_{\pm k}}(l)$. For $l$ large enough, but still smaller than $N_{r}$, the estimator $\bar{n}_{\pm k} (l)$ should be normally distributed with the expected mean and variance. The same procedure is applied to the estimators of $\bar{c}_{k, R}$ and $\bar{c}_{k, I}$.

\begin{figure}
\centering
\includegraphics[width=0.445\textwidth]{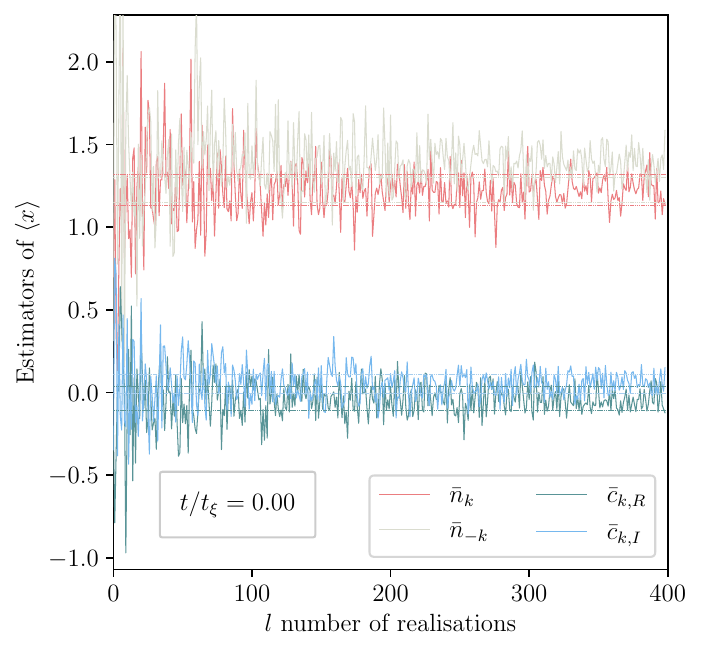}
\includegraphics[width=0.445\textwidth]{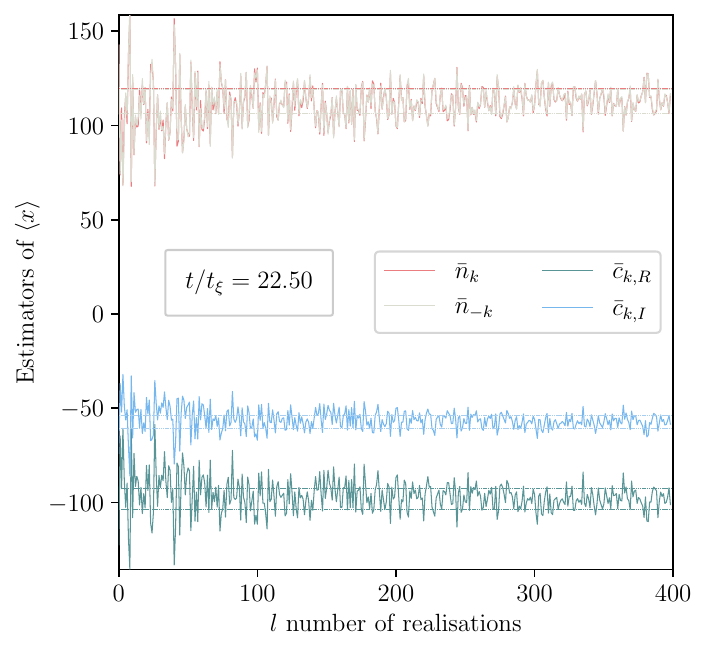}
\caption{Convergence of mean values $\langle x \rangle$ with number of realisations $l$. Full lines show the value of the estimators computed for a subsample of size $l$, picked with replacement from the full sample of $N_{r} =400$ realisations, as a function of $l$.
The dashed lines correspond to a window of size $\sqrt{S_{x}/N_{r}}$, where $S_{x}$ is given by Eq.~\eqref{def:estimator_variance}, on each side of the estimator $\bar{x}$ computed from the full set of $N_{r}$ realisations. For $\bar{n}$ this corresponds exactly to the range of values within the error bars shown in the figures of the main text. For both panels we used $k \xi = 1.0$ and $\rho \xi = 399.3$.
\label{fig:Convergence_estimators}}
\end{figure}

We perform two checks, showing first that the estimated mean values converge at the expected rate given by the CLT, and second that the resultant PDFs are well-approximated by normal distributions.
For the first check, we consider subsamples of increasing size $l$ and plot the corresponding estimators as a function of $l$. 
The resulting curves illustrate that the values of the estimators converge with increasing $l$, and that the residual fluctuations tend to occur within a window given by the estimated error, $\sqrt{S_{x}/N_{r}}$. This behaviour is entirely consistent with the expectations of the CLT.
The corresponding curves for some of the data used in Fig.~\ref{fig:nk_different_asap} are shown in Fig.~\ref{fig:Convergence_estimators}. 

\begin{figure}
\centering
\includegraphics[width=0.445\textwidth]{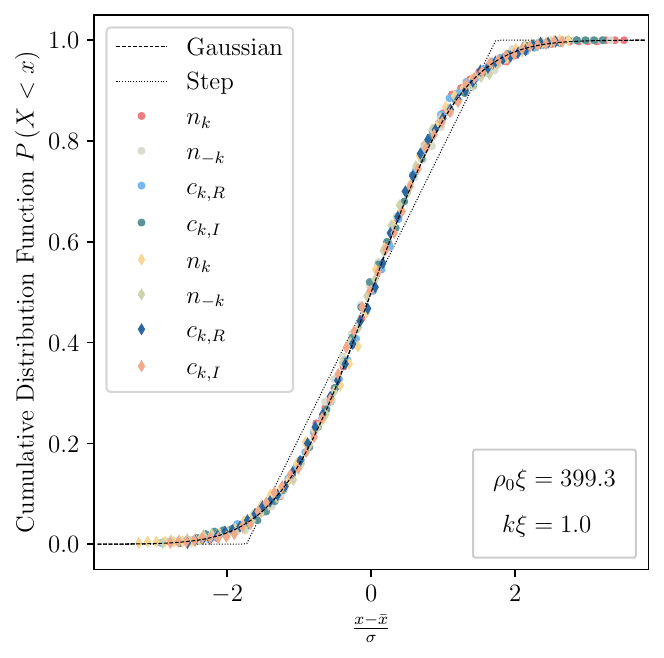}
\includegraphics[width=0.445\textwidth]{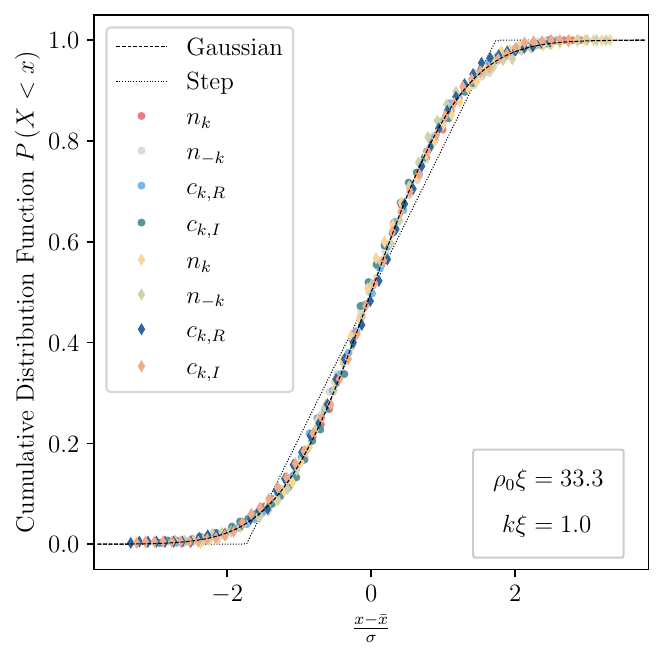}
\caption{CDF $\mathrm{P} ( X < x ) $ of the estimators $\bar{n}_k$, $\bar{n}_{-k}$, $\bar{c}_{k,R}$ and $\bar{c}_{k,I}$.
The coloured dots correspond to CDF computed from sample of $N_s$ estimators, e.g. $\{\bar{n}_k^{(j)} (l)  \}_{j \in [1,N_s]}$, bootstrapped from the samples of $N_{r} =400$ realisations used in Fig.~\ref{fig:nk_different_asap} at two times: initial time ($t/t_{\xi}=0$) represented by dots, and final time ($t/t_{\xi}=22.5$) represented by diamonds. They are compared to the CDF of a normal distribution (dashed lines) with the same mean and variance as that of the boostraped samples.
For the sake of comparison, we show in dotted line the CDF of a step distribution with normalised mean and variance. These plots were produced using $N_s = 400$ samples of size $l=200$ realisations picked with replacement out of the full set of $N_{r} =400$ realisations. The CDF was built using $100$ points equally spaced around $x = \bar{x}$.
\label{fig:1D_CDF}}
\end{figure}

Second, to check that $\bar{n}_{\pm k}$ is indeed normally distributed we again proceed by picking randomly $l$ out of our $N_{r}$ realisations (with replacement) and compute $\bar{n}_{\pm k} (l)$. 
We repeat the process $N_s$ times to obtain a collection $\{\bar{n}_{\pm k}^{(j)} (l)  \}_{j \in [1,N_s]}$ of $N_s$ values of the random variable $\bar{n}_{\pm k}(l)$ for a fixed $l$. We picked $l=200$ and $N_s =400$ for the figures. One way to check the normality of $\bar{n}_{\pm k} (l)$ would be to plot the histogram of the collection and check that it is well-described by a normal distribution. However, such a comparison is sensitive to the choice of binning for the histogram. Instead we build the cumulative distribution function (CDF) $\mathrm{P} ( \bar{n}_{\pm k} (l) < n ) $, which does not require any bin choice, and compare it to the CDF of a normal distribution with the same mean and variance as that computed from the collection $\{\bar{n}_{\pm k}^{(j)} (l)  \}_{j \in [1,N_s]}$. The same procedure is applied to $\bar{c}_{k, R}$ and $\bar{c}_{k, I}$. Some results are shown in Fig.~\ref{fig:1D_CDF}. The resulting CDFs match quite well that of a normal distribution, with mean and variance adjusted to those of the bootstrapped samples.

\subsection{Compound quantities not subject to CLT}

\subsubsection{Difficulties with compound quantities}

We have just shown that $\bar{n}_{\pm k}$, $\bar{c}_{k,R}$ and  $\bar{c}_{k,I}$, which are {\it bona fide} mean values built from a set of $N_{r}$ TWA simulations, obey Gaussian statistics to a very good approximation, {\it i.e.}, $N_{r}$ is large enough for the central limit theorem to apply.
However, in our analysis of the behaviour of the phonon state, we make use of quantities that are \textit{not} mean values.  In particular, we consider $\left| c_k \right|$, which is the norm of the mean value $c_k = c_{k,R} + i \, c_{k,I}$.  This is \textit{not} a {\it bona fide} mean value, and neither is the nonseparability parameter $\Delta_k = n_k - \left|c_k\right|$. (We recall that we defined this parameter using the number of excitations $n_{k}$ in the resonant mode with \textit{positive} wavenumber.)
They are estimated by $\left| \bar{c}_k \right| = \sqrt{ \bar{c}_{k, R}^2 + \bar{c}_{k, I}^2  }$ and 
$\bar{\Delta}_k = \bar{n}_{k} - \sqrt{ \bar{c}_{k, R}^2 + \bar{c}_{k, I}^2  }$.
These estimators will have their own PDF for a given number of realisations, but because they are not directly calculated as mean values, the CLT does not apply and we cannot assume {\it a priori} that their PDFs will become normal as $N_{r} \to \infty$. This is not necessarily an issue, but it behooves us to check how the typical PDFs of $\left|\bar{c}_k\right|$ and $\bar{\Delta}_k$ look to first give meaningful estimates for the error on these quantities, and then to validate our fit procedure, see Sec.~\ref{app:fitting}.

Let us first focus on the error bars. The ones given for $\left|\bar{c}_k\right|$ and $\bar{\Delta}_k$ in Figs~\ref{fig:Delta_different_asap}, \ref{fig:ck_different_A} and \ref{fig:Delta_different_A} come from estimates of their standard deviations. 
First, it is not always the case that the standard deviation is a good proxy for a confidence interval around the mean value, {\it e.g.}, for highly asymmetric PDFs a more sophisticated error analysis would be needed.
On the other hand, if we can check that the PDFs of $\left|\bar{c}_k\right|$ and $\bar{\Delta}_k$ are close to normal, this ensures that the window of two standard deviations around the mean correspond to a $68 \%$ interval. We do so numerically for different time points and parameters values.

Second, the covariances of the vector of basic quantities $\{ \bar{n}_k, \bar{n}_{-k} , \bar{c}_{k,R} , \bar{c}_{k,I} \}$ can be computed using the standard estimators generalising the ones for the variances
\begin{align}
\begin{split}
\label{def:estimator_cov}
S_{x,y} = \frac{1}{N_{r}} \sum_{i=1}^{N_{r}} \left( x^{(i)} - \bar{x} \right) \left( y^{(i)} - \bar{y} \right) \, ,
\end{split}
\end{align}
where $x$ and $y$ can either be $n_k$, $n_{-k}$, $c_{k,R}$ or $c_{k,I}$. However, the covariance of compound quantities, such as $\left|\bar{c}_k\right|$ and $\bar{\Delta}_k$ should be computed from their PDF inferred from that of $\{ \bar{n}_k , \bar{c}_{k,R} , \bar{c}_{k,I} \}$. It would be very costly to numerically compute such distribution for each data time point and parameter value. 
Instead, we estimate the covariances for compound quantities using the standard formulae for error propagation.
In general, for a vector of compound quantities $[f_1(n,c_{R},c_{I}),\ldots, f_N(n,c_{R},c_{I})]$, where $f_i$ is a differentiable function, the covariance matrix of the compound quantites $\Gamma_{f_1, \ldots , f_N}$ can be approximated by
\begin{equation}
\label{eq:propagation_cov}
\Gamma_{f_1, \ldots , f_N} =  J_{f_1,\ldots, f_N} . \Gamma_{n,c_{R},c_{I}} .  J_{f_1,\ldots, f_N}^{T} \, ,
\end{equation}
where 
\begin{equation}
J_{f_1,\ldots, f_N} = \begin{pmatrix}
\frac{\partial f_1}{\partial n} &  \frac{\partial f_1}{\partial c_R} &  \frac{\partial f_1}{\partial c_I} \\
& \ldots & \\
\frac{\partial f_N}{\partial n} &  \frac{\partial f_N}{\partial c_R} &  \frac{\partial f_N}{\partial c_I} 
\end{pmatrix} \, .
\end{equation}
For instance we have
\begin{align}
\label{eq:var_mod_c}
    \mathrm{Var} \left( \lvert c \rvert \right) 
    &= \left(\frac{\partial \lvert c \rvert}{\partial c_{R}}\right)^{2} \, {\rm Var}\left(c_{R}\right) 
    + \left(\frac{\partial \lvert c \rvert}{\partial c_{I}}\right)^{2} \, {\rm Var}\left(c_{I}\right) 
    + 2 \frac{\partial \lvert c \rvert}{\partial c_{R}} \, \frac{\partial \lvert c \rvert}{\partial c_{I}} \, {\rm Cov}\left(c_{R} , c_{I}\right) \\
    &= \frac{c^2_{\mathrm{R}} \, \mathrm{Var} \left( c_{\mathrm{R}} \right) + c^2_{I} \, \mathrm{Var}\left( c_{\mathrm{I}} \right) + 2 c_{\mathrm{R}} c_{\mathrm{I}} \, \mathrm{Cov} \left( c_{\mathrm{R}} , c_{\mathrm{I}}\right) }{ c^2_{R} + c^2_{I}} \, .
\end{align}
An additional check would be to compare the resulting covariance matrix elements to that computed directly from the numerically computed full PDF.
We combine this check with the normality check: for given means $\{ \langle n_k \rangle , \langle n_{-k} \rangle , \langle c_{k,R} \rangle , \langle c_{k,I} \rangle \}$ and covariance $\Gamma_{n_k n_{-k} c_R c_I}$ for $\{ \bar{n}_k , \bar{c}_{k,R} , \bar{c}_{k,I} \}$, we directly compare the numerically computed full PDF for compound quantities ({e.g. $|\bar{c}_k|$) and a normal distribution supplied with covariance approximated by $\Gamma_{n_k n_{-k} c_R c_I}$ using Eq.~\eqref{eq:propagation_cov} ({\it e.g.}, $\mathcal{N} [ \sqrt{ \bar{c}_{k, R}^2 + \bar{c}_{k, I}^2  } , \mathrm{Var} ( | \bar{c}_k | ) ]$ where the estimators are computed over $N_{r}$ realisations and the variance is given by Eq.~\eqref{eq:var_mod_c}).
In the next section we only discuss the diagonal terms; see App.~\ref{app:joint_normality} for discussion of the off-diagonal terms, which characterise the joint probability distribution functions.

\subsubsection{Normality checks for compound quantities}
\label{app:normal_compound}

Given that $| \bar{c}_k |$ and $\bar{\Delta}_k$ are directly related to $\bar{n}_k$, $\bar{c}_{k,R}$ and $\bar{c}_{k,I}$, we can easily construct their PDFs for any joint Gaussian PDF of $\{ \bar{n}_k , \bar{c}_{k,R} , \bar{c}_{k,I} \}$. In the rest of this section we focus on $| \bar{c}_k |$, but the exact same procedure is applied to $| \bar{\Delta}_k |$. Note that we do not need to include $n_{-k}$ in the picture given that we define $\Delta_k$ using $n_k$ only.

We first construct the joint PDF of $\bar{n}_k$ and $\left|\bar{c}_k\right|$ by integrating over the phase of $\bar{c}_k = \bar{c}_{k,R} + i \, \bar{c}_{k,I}$:
\begin{align}
\begin{split}
\label{def:PDF_n_modc}
P\left(\bar{n}_k = n, \lvert \bar{c}_k \rvert = \left|c\right|\right) & = \int_{- \infty}^{+ \infty} \int_{- \infty}^{+ \infty} \delta \left( \sqrt{c_{R}^2 + c_i^2} - \lvert c \rvert \right)  P\left(n, c_{R} , c_{I} \right) \mathrm{d} c_{R} \mathrm{d} c_{I} \, , \\
& = \int_{0}^{ 2 \pi}  \lvert c \rvert P\left(n, \lvert c \rvert \cos \theta , \lvert c \rvert \sin \theta \right)  \mathrm{d} \theta \, ,
\end{split}
\end{align}
where from the first to the second line we performed the change of variable $(c_r, c_I) \to (|c| , \theta = \mathrm{Arg} [c])$. We can then further trace over the values of $\bar{n}_k$ to get the PDF of $| \bar{c}_k |$
\begin{equation}
\label{eq:PDF_mod_c}
P\left( \lvert \bar{c}_k \rvert = \left|c\right|\right) = \int_{- \infty}^{+ \infty} P\left(\bar{n}_k = n, \lvert \bar{c}_k \rvert = \left|c\right|\right)  \mathrm{d} n \, .
\end{equation}
We want to compare this PDF to a normal distribution.
Since they are \textit{a priori} quite different for generic values of $\{ \langle n_k \rangle , \langle c_{k,R} \rangle , \langle c_{k,I} \rangle \}$ and covariance $\Gamma_{n c_R c_I}$, to make our comparison relevant we choose them to match that obtained in TWA simulations.
The means and covariance are estimated in TWA simulations using the estimators $\{ \bar{n}_k, \bar{c}_{k,R}, \bar{c}_{k,I} \}$ and $\{ S_{x,y} \}_{x,y = n_k, c_{k,R}, c_{k,I}}$ computed over $N_{r}$ realisations for some time point and values of the parameters. We then numerically compute the joint PDF using Eq.~\eqref{eq:PDF_mod_c}~\footnote{Notice that the mean value of $|\bar{c}_{k}|$ computed from this PDF by
\begin{equation}
    \mathbb{E} \left[ \lvert \bar{c}_{k} \rvert \right] = \int_{0}^{+ \infty} \left|c\right| P\left( \lvert \bar{c}_k \rvert = \left|c\right| ; \bar{n}_k, \bar{c}_{k,R}, \bar{c}_{k,I} , S_{x,y}  \right) \mathrm{d}  \left|c\right| \, ,
\end{equation}
will differ from the value computed as $\sqrt{ \bar{c}_{k,R}^2 + \bar{c}_{k,I}^2 \label{footnote:naive_means}}$, which is the one used in the figures of the paper. This is the usual difference between applying a function to the mean value of a random variable, and taking the mean value of a random variable on which the function was applied. This discrepancy should be small for large $N_{r}$ since the distributions of $\bar{c}_{k,R/I}$ will then be very peaked around their mean values.}. 
Finally, we compare this PDF to $\mathcal{N} [ \sqrt{ \bar{c}_{k, R}^2 + \bar{c}_{k, I}^2  } , \mathrm{Var} ( | \bar{c}_k | ) ]$ using the same values of the estimators of means and covariance and Eq.~\eqref{eq:var_mod_c}.

In Fig.~\ref{fig:PDF_modc_Delta} we show the results of this comparison for different values of means and covariance matrix. The results for the very same procedure applied to $\bar{\Delta}_k$ are shown in the same figure. A couple of comments are in order.
First, we observe a significant mismatch between the distributions of $|\bar{c}_k|$ and  $\bar{\Delta}_k$ at $t / t_{\xi} = 0$. This behaviour is in fact confined to very early times, which can be explained in the following way.
Since $|\bar{c}_k|$ is by construction positive, there is a sharp drop in the actual distribution before $|\bar{c}_k| = 0$, as seen in the red dots in Fig.~\ref{fig:PDF_modc_Delta}.
By contrast, if we were to assume that $|\bar{c}_k|$ follows a normal distribution, then $|\bar{c}_k|$ would take negative values whenever its mean and standard deviation are of the same order. This typically happens at very early times ($t / t_{\xi} = 0$ in Fig.~\ref{fig:PDF_modc_Delta}), and we thus conclude that the probability distribution of $|\bar{c}_k|$ cannot be normal there.
However, for later times ($t / t_{\xi} = 12$ and $t / t_{\xi} = 17$ in Fig.~\ref{fig:PDF_modc_Delta}) the mean of $|\bar{c}_k|$ is large enough to prevent this behaviour.
Second, the agreement would have been further improved had we used a normal distribution with mean and variance computed from the exact PDF of $|\bar{c}_k|$ and $\bar{\Delta}_k$, as in Eq.~\eqref{eq:PDF_mod_c}, rather than the values given by the estimators for $N_r =400$ (see footnote~\ref{footnote:naive_means}). Therefore, part of the disagreement at early time is due to the choice of parameters characterising the reference normal distribution rather than a deviation from normality in the exact PDF.
That being said, the distributions of $|\bar{c}_k|$ and $\bar{\Delta}_k$ are generally in very good agreement with the reference normal distributions.
This justifies using the square root of the variance computed from error propagation as an indication of the error for these two quantities: roughly $68\%$ of values should be in this range.

\begin{figure}
\centering
\includegraphics[width=0.405\textwidth]{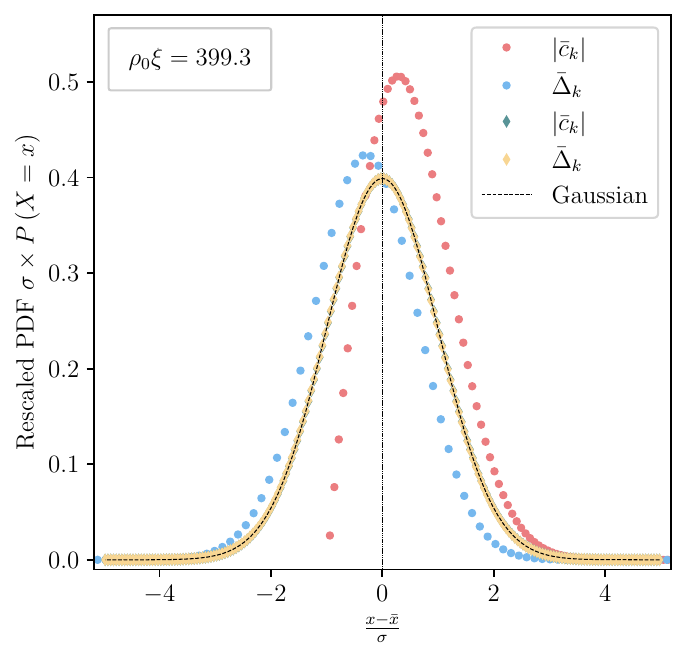}
\includegraphics[width=0.405\textwidth]{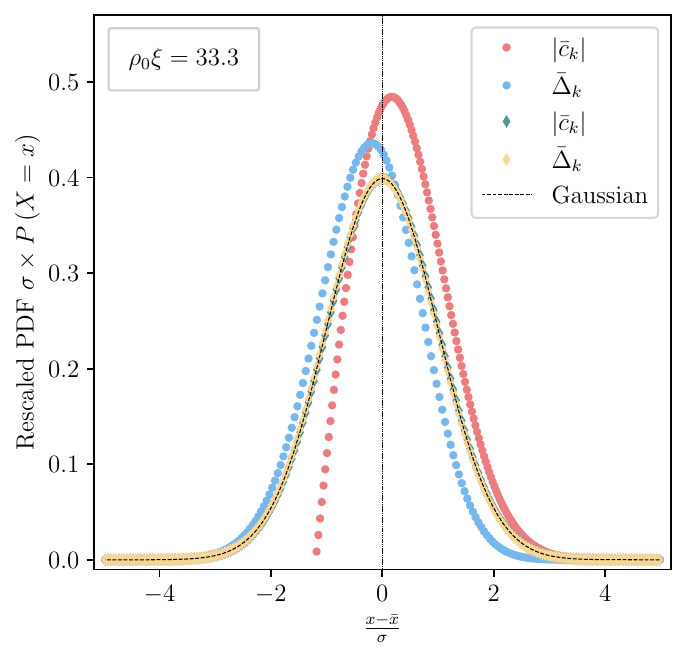}
\caption{
PDF of $|\bar{c}_k|$ and $\bar{\Delta}_k$ for normally distributed $\{\bar{n}_k, \bar{c}_{k,R} , \bar{c}_{k,I} \}$ with means and covariances numerically extracted from TWA simulations with resonant modes $k \xi = \pm 1.0$ for two different values of the gas density $\rho_0 \xi$. The dots (respectively diamonds) show the PDF corresponding to means and covariances at $t /t _{\xi} = 0$ (resp.  $t /t _{\xi} = 22.5$ ). Note that we normalised the numerically computed PDF using our \textit{estimators} of the means and variances computed over the whole set of $N_r = 400$ realisations, which might differ from the actual means and covariances of the PDF (see footnote~\ref{footnote:naive_means}). This might result in non-centred, non-unit variance PDFs when these numbers significantly differ, as is the case for $t /t _{\xi} = 0$.
The other parameters of the simulations are reported in Tab.~\ref{tab:fixed_param}
\label{fig:PDF_modc_Delta}. }
\end{figure}

\subsection{Fitting procedure and errors}
\label{app:fitting}

In Fig.~\ref{fig:fit_n_c_different_asap_early_times} we showed the results of fitting the growth of $n_k$ and $\lvert c_k \rvert$ to the functional forms derived in Eqs.~\eqref{eq:n_c_initial_thermal}. 
The best-fit value $\alpha^{\star}$ for the decay to growth rate ratio $\alpha$ is found using a least-$\chi^{2}$ approach, {\it i.e.}, by minimising w.r.t. $\alpha$ the sum over all time points $t_{i}$ of the squared differences between the numerically obtained values of $n_{\mathrm{TWA}}$ and $\lvert c \rvert_{\mathrm{TWA}}$ 
and the predicted values $n_{\mathrm{pred}}$ and $\lvert c \rvert_{\mathrm{pred}}$ given by Eqs.~\eqref{eq:n_c_initial_thermal}. 
The differences are weighted by the estimated covariance matrix at each time point. Mathematically, the quantity minimised over $\alpha$ is 
\begin{equation}
\label{def:distance_fit}
d\left( \alpha \right) = \sum_{i=1}^{n_t} \delta V^{T} \left( t_i ; \alpha \right) . \Gamma^{-1}_{n \lvert c \rvert} \left( t_i \right) . \delta V \left( t_i ; \alpha \right) \, ,
\end{equation}
where $n_t$ is the number of time points $t_i$ sampled in the simulations,  $\delta V ( t ; \alpha ) = [  n_{\mathrm{TWA}} (t) - n_{\mathrm{pred}} (t ; \alpha ) , |c|_{\mathrm{TWA}} (t) - |c|_{\mathrm{pred}} (t ; \alpha ) ]^T$ and the covariance matrix $\Gamma_{n \lvert c \rvert} \left( t \right)$ is approximated using Eq.~\eqref{eq:propagation_cov}, based on $\Gamma_{n,c_{R},c_{I}}$ estimated from the TWA simulations using Eq.~\eqref{def:estimator_cov}.
If the errors given by $\delta V ( t_i ; \alpha_{\star} )$ are normally distributed with covariance matrix $\Gamma_{n \lvert c \rvert} \left( t_i \right)$ and vanishing mean (meaning that our model perfectly describes the observed average values when $\alpha = \alpha^{\star}$), then by diagonalising $\Gamma_{n \lvert c \rvert} \left( t_i \right)$ one can check that $\delta V^{T} \left( t_i ; \alpha_{\star} \right) . \Gamma^{-1}_{n \lvert c \rvert} \left( t_i \right) . \delta V \left( t_i ; \alpha_{\star} \right)$ is a sum of two independent centred normal variables with unit variance\footnote{Strictly speaking we should thus check that $\bar{n}_k$ and $|\bar{c}_k|$ are \textit{jointly}, not just separately, normally distributed. Some partial checks are presented in App.~\ref{app:joint_normality}.}. $d (\alpha)$ is then the sum of $2 n_t$ independent centred normal variables with unit variance and thus follows a $\chi^2 ( 2 n_t )$ distribution. 
A $\chi^2 ( N )$ distribution has mean value $N$ and variance $2 N$. For large enough $N$ it is therefore quite peaked. A usual goodness-of-fit quantifier is then to check that for the best-fit value $\alpha_{\star}$, we have $d ( \alpha_{\star} ) \approx N$. The error on the best-fit value $\alpha_{\star}$ is then estimated by looking for the values of $\alpha$ where $d ( \alpha ) = d ( \alpha_{\star} ) + N$, or equivalently $d ( \alpha ) / N = d ( \alpha_{\star} )/N + 1$ where $d ( \alpha ) / N$ is known as the reduced chi-squared.
(Note that in fact the number of independent degrees of freedom is only $2 n_t-1$, because one degree of freedom is lost in determining $\alpha_{\star}$ which relates the different time points.) To quantify the extent to which our theoretical prediction deviates from the observations, we give in Table~\ref{tab:fig5_parameters} the values of the reduced chi-squared $\chi_{\nu}^{2} = d\left(\alpha_{\star}\right)/(2 n_t-1)$.  If our model were a good fit we would expect $\chi_{\nu}^{2} \sim 1$, whereas $\chi_{\nu}^{2}$ significantly larger than $1$ indicates that the model does not describe the data all that well.

This procedure is applied to obtain the error bars in Fig.~\ref{fig:fit_n_c_different_asap_early_times}. The resulting error bars are so small as not to be visible in the plot. Why is that? The reason is that the above procedure for error estimation somehow assumes that the model used to describe the data is \textit{exact}, and that discrepancies between the predicted values and that obtained from the TWA simulations are due to statistical variance. The resulting uncertainty on the fitted parameter has to be understood as the uncertainty on the parameter \textit{were the model exact}.  However, if the model is inaccurate (and it always is to a certain extent), this inaccuracy is not captured by the above procedure. For concreteness, consider the curve corresponding to $\alpha =0.82$ in Fig.~\ref{fig:nk_different_asap}. We have $n_t =281$ data points with relatively small uncertainties. Given the model, the curve corresponding to the parameter value minimising the residuals is shown in solid line, and is obviously not a good description of the evolution of $n_k$. Yet, any small variation in the value of $\alpha$ would drastically increase the sum of residuals given that the error bars are small and the number of data points is large. In other words: the uncertainties on the data points are small enough so that there is a very well-defined notion of best description within the class of models we allow for, but no such model is an accurate description of the evolution. This discrepancy is not captured by the uncertainty over $\alpha$ given by the fitting procedure.

\subsection{Joint normality checks}
\label{app:joint_normality}

In this final appendix we present further checks of our error estimation procedures.
We have shown in the previous appendices that the number of realisations $N_r$ used in the simulations was sufficient for it not to be unreasonable to assume that the estimators for the dubbed "basic" quantities $\bar{n}_{k}$, $\bar{n}_{-k}$, $\bar{c}_{k, R}$ and $\bar{c}_{k, I}$ are \textit{separately} normally distributed with the expected mean and variance i.e. that $N_r$ is large enough so that the CLT can be applied to these estimators.
However, in our analysis we also considered compound quantities derived by combining these different estimators which do not sastify the CLT, such as $\Delta_k = n_k - |c_k|$ which is approximated by $\bar{\Delta}_k = \bar{n}_{k} - \sqrt{ \bar{c}_{k, R}^2 + \bar{c}_{k, I}^2  }$. 
We also wish to check that the {\it joint} distribution of the basic quantities $(\bar{n}_{k} , \bar{n}_{-k} , \bar{c}_{k, R} , \bar{c}_{k, I} )$ can also be well approximated as normal. This is expected from the CLT and we confirm it in the first part of this section. Note that the four variables are expected to be correlated, being built out of the same underlying realisations $\{ b_k^{(i)} \}_i$, so the previous checks of separate normality are indeed not sufficient to conclude joint normality.
To simplify the analysis we will assume complete isotropy of the realisations, {\it i.e.}, that $n_k^{(i)} = n_{-k}^{(i)}$ for any realisation $i$, even though the TWA data is only statistically isotropic with $n_k^{(i)}$ and $n_{-k}^{(i)}$ being independent realisations of the same probability distribution. Thus, instead of having to study the vector of four quantities $\{ \bar{n}_k,  \bar{n}_{-k} , \bar{c}_{k,R} , \bar{c}_{k,I} \}$, we restrict our study to three of them, $\{ \bar{n}_k, \bar{c}_{k,R} , \bar{c}_{k,I} \}$. Note again that this is sufficient if we only want to describe the compound quantities $|c_k|$ and $\Delta_k$ that do not involve $n_{-k}$.
Then we discuss whether the estimator of $|c_k|$, which has been shown in previous sections to be approximately normally distributed, is also \textit{jointly} normally distributed with the estimator of $n_k$. This joint normality assumption is implictly used to fully justify the error bars on the fitted parameter $\alpha$, see App.~\ref{app:fitting}.

\subsubsection{Joint normality approximation checks for basic quantities}

First, we have to estimate the off-diagonal elements of the covariance matrix: $\mathrm{Cov} (\bar{n}_k, \bar{c}_{k,R})$, $\mathrm{Cov} (\bar{n}_k, \bar{c}_{k,I})$ and $\mathrm{Cov} ( \bar{c}_{k,R}, \bar{c}_{k,I})$. We do so by using the estimators~\eqref{app:error}.
Next, using a similar bootstrapping procedure as described in App.~\ref{app:normal_basic}, we check that the CDF of \textit{pairs}  $(\bar{n}_{k} , \bar{c}_{k, R} )$, $(\bar{n}_{k} , \bar{c}_{k, I} )$ and $( \bar{c}_{k, R} , \bar{c}_{k, I} )$ are in good agreement with those of normal distributions with the same parameters. Obviously these checks are only partial since we should further check that three-dimensional joint CDF of the vector $\{ \bar{n}_k , \bar{c}_{k,R} , \bar{c}_{k,I} \}$ is also normal. Such comparison cannot be done visually anymore and would require a more complicated quantitative estimation with the expected normal CDF. For this reason we limit ourselves to the above checks. 

We pick randomly $l$ out of our $N_{r}$ realisations with replacement and compute from these the three estimators $\bar{n}_k$, $\bar{c}_{k,R}$ and $\bar{c}_{k,I}$. We repeat the process $N_s$ times for a fixed value of $l$ and obtain a collection of vectors of estimators $\{ [\bar{n}_k^{(j)} (l) , \bar{c}_{k,R}^{(j)} (l)  , \bar{c}_{k,I}^{(j)} (l)  ]  \}_{j \in [1,N_s]}$.
From these values we build the joint two-dimensional CDF for each pair, {\it e.g.}, $\mathrm{P} ( \bar{n}_k (l) < n , \bar{c}_{k,R} (l) < c_R ) $. We want to compare this CDF to that of a 2D normal distribution with the same means and covariance matrix as $\{ \bar{n}_k (l) , \bar{c}_{k,R} \}$ in the boostrapped sample, $\mathrm{P}_{\mathrm{norm.}} ( \bar{n}_k (l) < n , \bar{c}_{k,R} (l) < c_R)$.
A visual comparison is harder than for the one-dimensional case. We therefore plot the difference between the inferred and reference CDFs
\begin{equation}
\label{def:deltaCDF}
\delta CDF ( \bar{x} , \bar{y} ) = \mathrm{P} ( \bar{x} < x , \bar{y}  < y ) - \mathrm{P}_{\mathrm{norm.}} ( \bar{x} < x , \bar{y}  < y ) \, .
\end{equation}
For instance for $\bar{n}_k$ and $\bar{c}_{k,R}$ we get $\delta CDF ( \bar{n}_k , \bar{c}_{k,R} ) = \mathrm{P} ( \bar{n}_k (l) < n , \bar{c}_{k,R} (l) < c_R ) - \mathrm{P}_{\mathrm{norm.}} ( \bar{n}_k (l) < n , \bar{c}_{k,R} (l) < c_R)$.
By construction $\delta CDF ( \bar{x} , \bar{y} ) \in [-1,1]$ so that a difference of $0.1$ is already substantial. 
The results for some data of Fig.~\ref{fig:1D_CDF} are shown in Fig.~\ref{fig:2D_CDF}. Different values of the parameters would lead to similar plots. In general the differences between the approximated CDF and that of the reference normal distribution are small, typically not more than $0.05$, and the regions with discrepancies of this order are quite restricted so that the normality approximation is very good overall.

\begin{figure}
\centering
\includegraphics[width=0.33\textwidth]{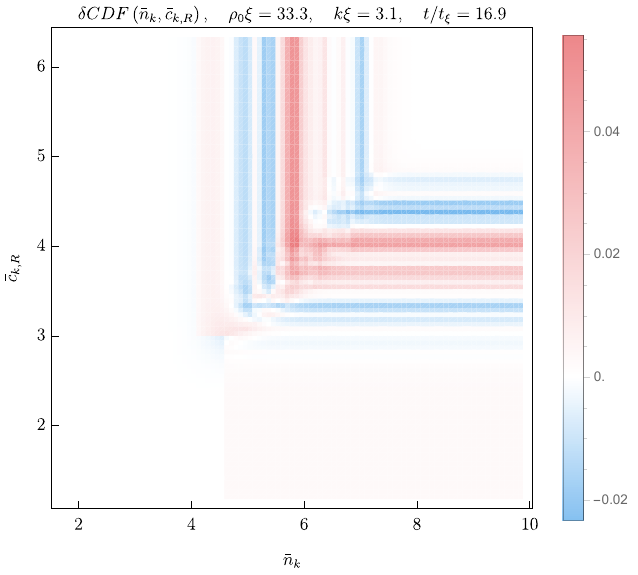}
\includegraphics[width=0.33\textwidth]{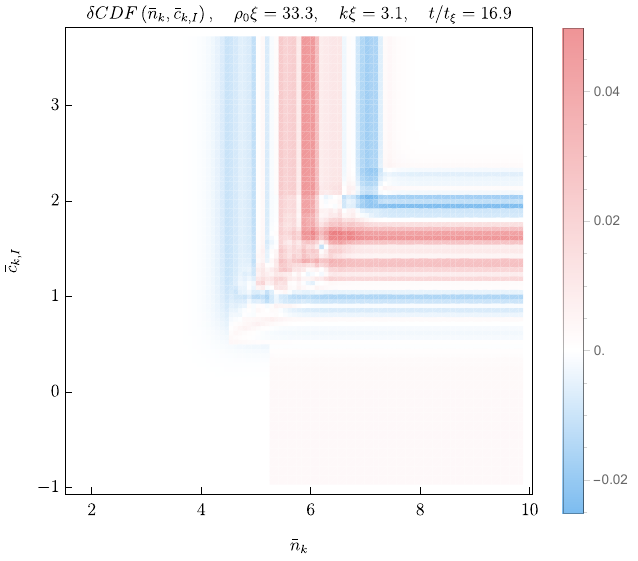}
\includegraphics[width=0.33\textwidth]{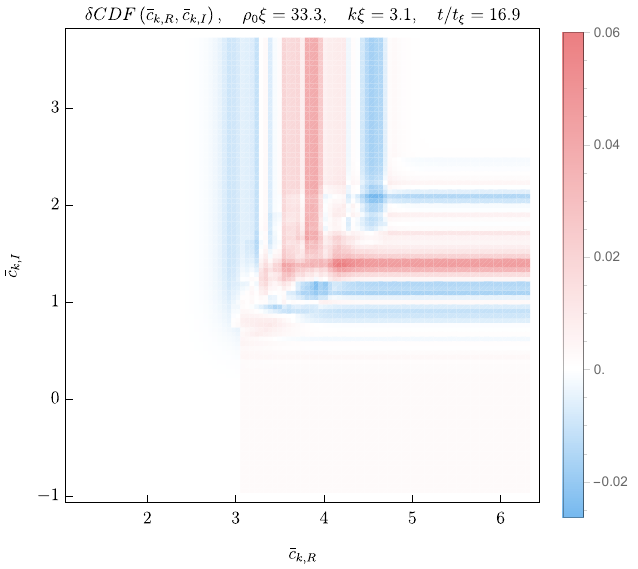}
\includegraphics[width=0.33\textwidth]{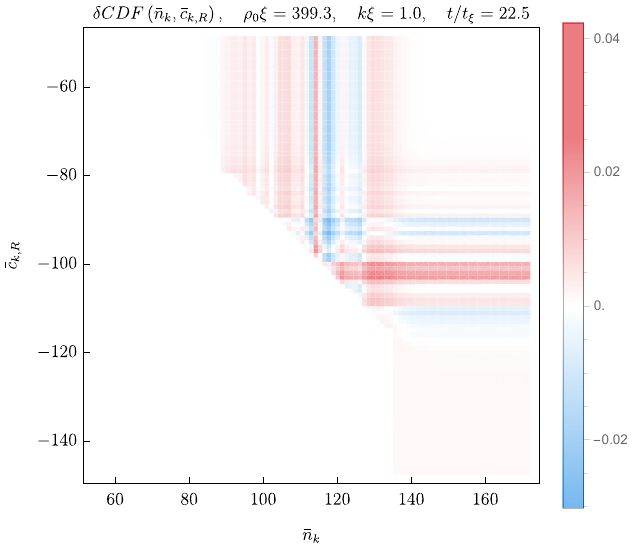}
\includegraphics[width=0.33\textwidth]{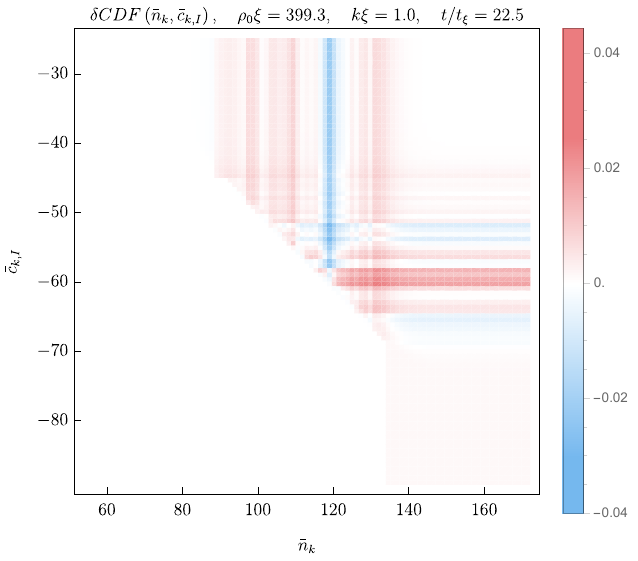}
\includegraphics[width=0.33\textwidth]{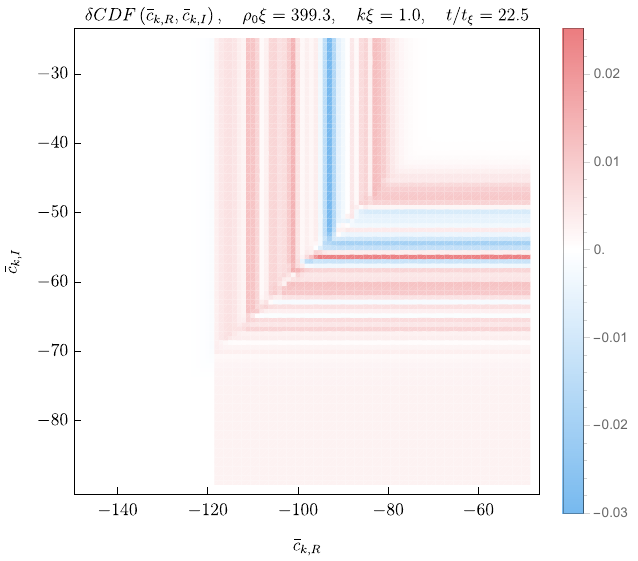}

\caption{Difference $\delta CDF$ defined in Eq.~\eqref{def:deltaCDF} between 2D CDFs for any pair of estimators as estimated from TWA data, {\it e.g.}, $\mathrm{P} ( \bar{n}_k (l) < n , \bar{c}_{k,R} (l) < c_R ) $ for $\{ \bar{n}_k , \bar{c}_{k,R} \}$, and the CDF of a normal distribution with the same means and covariances, {\it e.g.}, $\mathrm{P}_{\mathrm{norm.}} ( \bar{n}_k (l) < n , \bar{c}_{k,R} (l) < c_R)$. 
Note that the color bar is picked with fixed reference red and blue colors for the maximum and minimum value respectively. When one color seems absent from the color bar legend it is because the value it is coding for is too close to $0$.}
\label{fig:2D_CDF}
\end{figure}

\subsubsection{Joint normality check for $\bar{n}_k$ and $|\bar{c}_k|$}

The normality assumption, checked separately for $|\bar{c}_k|$ and $\bar{\Delta}_k$ in Sec.~\ref{app:normal_compound}, is also necessary to justify the error estimation in the fitting procedure.
The assumption that the quantity $d ( \alpha )$ defined in Eq.~\eqref{def:distance_fit} follows a $\chi^2$ distribution is only justified when the error vector $\delta V$ is normally distributed, 
which requires again a \textit{joint} normality check for the fitted quantities, here $n_k$ and $|\bar{c}_k|$. 
To simplify the checks, we first assume that indeed our model is at any time a perfect description of the mean value of the distribution of $\bar{n}_k$ and $|\bar{c}_k|$ e.g. $\langle n_{\mathrm{TWA}} (t) \rangle = n_{\mathrm{mod.}} (t ; \alpha )$. Note that this is slightly incorrect as can be seen by the inaccuracy of the best fit obtained for instance for the curve corresponding to $\alpha =0.82$ in Fig.~\ref{fig:nk_different_asap}.
We then have to justify that $ [  n_{\mathrm{TWA}} (t) , |c|_{\mathrm{TWA}} (t)  ]^T$ is normally distributed about this mean. Using the same strategy as in Sec.~\ref{app:normal_compound}, for given values of parameters and time point, we estimate the means $\{ \langle n_k \rangle , \langle c_{k,R} \rangle , \langle c_{k,I} \rangle \}$ and covariance $\Gamma_{n c_R c_I}$ of $\{ \bar{n}_k , \bar{c}_{k,R} , \bar{c}_{k,I} \}$ in the TWA simulations using the usual estimators. We first compute the joint PDF of $\bar{n}_k$ and $| \bar{c}_k |$ using Eq.~\eqref{def:PDF_n_modc}, and then their joint CDF. 
Next, using Eq.~\eqref{eq:propagation_cov} we get from $\Gamma_{n c_R c_I}$ an approximate of $\Gamma_{n |c|}$ the covariance of $\bar{n}_k$ and $| \bar{c}_k |$, and we compute the 2D normal distribution $\mathcal{N} [ \bar{n}_k, \sqrt{ \bar{c}_{k, R}^2 + \bar{c}_{k, I}^2  } , \Gamma_{n |c|} ]$, and its CDF. Finally, we take the difference as in Eq.~\eqref{def:deltaCDF}. 
The results are shown in Fig.~\ref{fig:2D_CDF_n_mod_c}.
As for the 1D PDFs, the largest deviations from normality are observed at early times and the agreement is very good at late times.

\begin{figure}
\centering
\includegraphics[width=0.405\textwidth]{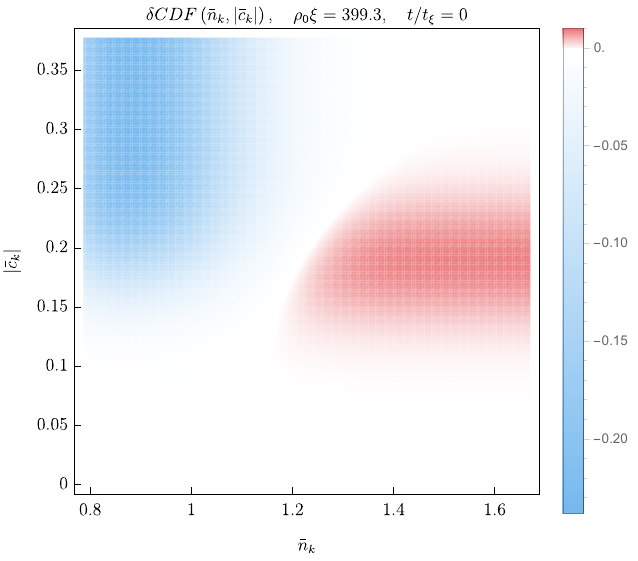}
\includegraphics[width=0.405\textwidth]{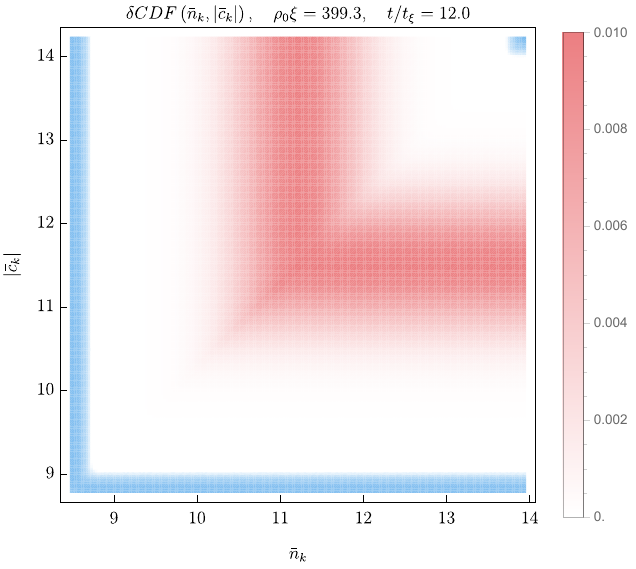}
\includegraphics[width=0.405\textwidth]{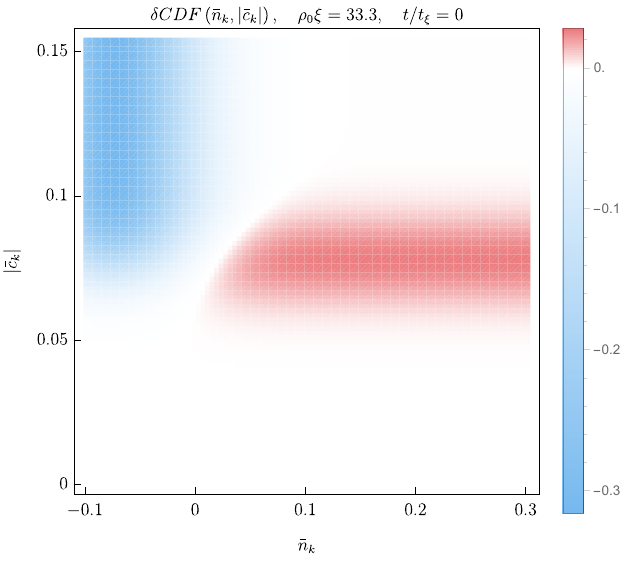}
\includegraphics[width=0.405\textwidth]{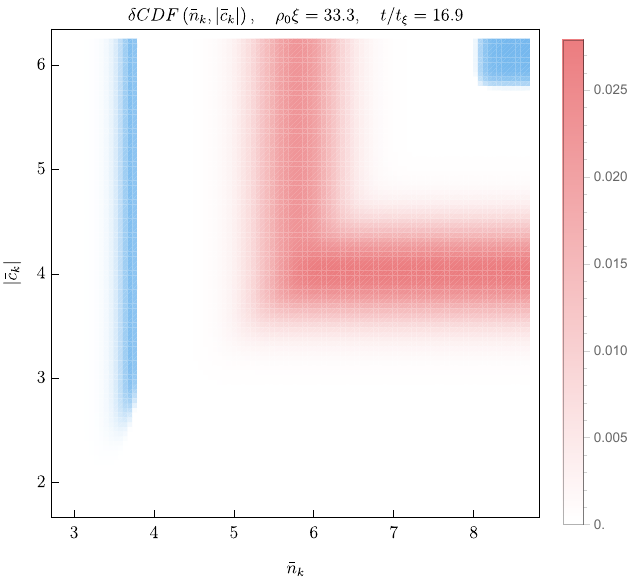}
\caption{Difference $\delta CDF$ defined in Eq.~\eqref{def:deltaCDF} between $\mathrm{P} ( \bar{n}_k < n , |\bar{c}_k| < |\bar{c}| ) $ for $\{ \bar{n}_k , |\bar{c}_k| \}$, the two-dimensional CDF for $ \bar{n}_k$ and $|\bar{c}_k|$ numerically computed from Eq.~\eqref{def:PDF_n_modc}, and $\mathrm{P}_{\mathrm{norm.}}  ( \bar{n}_k < n , |\bar{c}_k| < |\bar{c}| ) $, the CDF of a normal distribution with the same means and covariance approximated using Eq.~\eqref{eq:propagation_cov}. The means $\langle n \rangle, \langle c_{R} \rangle, \langle c_I \rangle$ and covariance $\Gamma_{n c_R c_I}$ are extracted from TWA simulations for the resonant modes with parameter values of parameters and times in the plot. The upper panels correspond to $k \xi = \pm 1.0$ and the lower panels to $k \xi = \pm 3.1$. 
Note that the color bar is picked with fixed reference red and blue colors for the maximum and minimum value respectively. When one color seems absent from the color bar legend it is because the value it is coding for is too close to $0$.}
\label{fig:2D_CDF_n_mod_c}
\end{figure}

\bibliographystyle{crunsrt}

\nocite{*}

\bibliography{samplebib}

\end{document}